\documentclass[prc,showpacs,showkeys,superscriptaddress,nofootinbib,floatfix,twocolumn]{revtex4}
\usepackage{color}
\usepackage{amsfonts}
\usepackage{amsbsy}
\usepackage{mathrsfs}
\usepackage{graphicx}
\usepackage{subfigure}
\newcommand{\beq}{\begin{equation}}
\newcommand{\eeq}{\end{equation}}
\newcommand{\bea}{\begin{eqnarray}}
\newcommand{\eea}{\end{eqnarray}} 
\begin{document}
	%
\title{Screening properties of quark-gluon plasma obtained from distribution and correlation functions of the constituent quasiparticle model}

\author{V.~Filinov}
\thanks{Corresponding author\quad E-mail:~\textsf{vladimir\_filinov@mail.ru}}
\author{A.~Larkin}
\author{V.~Fortov}
\affiliation{Joint Institute for High Temperatures, Russian Academy of Sciences, Izhorskaya 13, Bld. 2, 125412 Moscow, Russia }
\date{\today}

\begin{abstract}
Based on the constituent quasiparticle model of quark-gluon plasma (QGP), the matrix elements of the density operator and the Wigner function in the color phase space are presented in form of color path integrals over Wiener and SU(3) group Haar measures.
Monte Carlo calculations of quark and gluon momentum distributions and spatial pair distribution functions have been carried out for the strongly coupled QGP plasma in thermal equilibrium at zero baryon chemical potential.
The Debye screening mass and the running coupling constant have been obtained from the spatial pair distribution function and are in agreement with the available lattice QCD data.
At densities related to the average interparticle distance more than 0.4~fm the gluon bound states in the form of glueballs have been found.
Comparison with the Maxwell – Boltzmann distribution shows significant influence of interparticle interaction on high energy asymptotics of the momentum distribution functions, resulting in appearance of quantum “tails”.
The new color pair correlation function has been introduced, and related new color screening mass has been discussed.
 \end{abstract}

\pacs{12.38.Mh, 12.38.Lg, 25.75.Nq, 25.75.Dw}  
\keywords{quark-gluon plasma, Wigner function, screening properties, momentum distribution, glueballs}

\maketitle

\section{Introduction} \label{s:intro}  
The color screening of heavy quarks, as it was proposed in \cite{Matsui}, should lead to dissociation of  $J/\psi$ bound state and may indicate a formation of the quark-gluon plasma (QGP) arising in heavy-ion collision experiments \cite{Laine}.
The idea of a screening mass and running coupling constant allow us to understand the interaction of a particles, inserted into a medium, intuitively.
The simplest way is to consider a static quark-antiquark probe in color plasma, wherein all medium effects are taken into account by two body interaction potential $U(r)$.
In medium both the Coulomb and string-like parts of interaction between the heavy quark and anti-quark are modified \cite{Koma,Kaczmarek,Burnier}, and the potential $U(r)$ is generally a complex quantity \cite{Beraudo,Brambilla1,Brambilla2}.
An effective description have to capture the effects of screening and Landau damping, related to the real and imaginary parts of the potential correspondingly.
Several approaches were proposed to treat this problem using e.g. effective theories 
\cite{Kajantie,Hart,Philipsen}, spatial correlation functions 
or the behavior of the color singlet free energies \cite{Nadkarni1,Nadkarni2,Maezawa,Digal}.   
However the perturbative expansion in terms of the QCD coupling constant $g$ fails \cite{Linde}.  
Therefore it is necessary to use a non-perturbative approaches to make prediction of QGP properties.
Currently, the lattice QCD simulation is 
the only systematic method allowing to extract various properties 
of QGP and to study the inter-quark interactions and screening properties of QGP. 
However interpretation of these very complicated numerical computations requires
application of various QCD motivated, albeit schematic, models simulating
various aspects of the full theory and allowing for a deeper physical
understanding. 
Moreover, such models are needed in cases when the lattice QCD
fails, e.g. at large quark chemical potentials and out of thermodynamic equilibrium. 
For temperatures higher than the QCD transition temperature  ( of order $175$~MeV) this issue can be addressed 
by adopting the quasiparticle approaches 
\cite{Chandra1,Chandra2,Chandra3,Peshier1,Peshier2,Peshier3,Peshier4,Dumitru,Fukushima,Ghosh,Abuki,Tsai,
Ruggieri,DElia1,DElia2,Castorina1,Castorina2,Plumari}. 
quasiparticle models have to account such high temperatures for the hot QCD medium effects and to describe the influence 
of strong inter-particle interaction. 
To consider here the microphysics insights into the inter-quark interactions and screening mechanism, we investigate QGP in framework of quasiparticle QGP model, which has been proved in reproducing various aspects of QGP thermodynamics and kinetic properties \cite{QGP1,QGP2,EbelForFil}.
The approach used in \cite{QGP1,QGP2,EbelForFil} is based on the quasiparticle pattern and is motivated by expectation that the main features of non-Abelian plasmas can be understood in simple semi-classical terms without difficulties inherent to a full quantum field-theoretical analysis.
This also sets the motivation for the investigations, which are using the continuous classical color variable interacting with the chromodynamic field \cite{LM105,Mr,reveiw,shuryak1}.
Quantum Monte Carlo simulations presented in \cite{EbelForFil} 
were able to reproduce the lattice equation of state even near the critical temperature and  
at nonzero baryon chemical potential. 

The quasiparticle model of QGP is seen to be consistent with the lattice simulations  \cite{Boyd1,Boyd2,Panero,Karsch,Cheng1,Cheng2,Bazavov,Borsanyi1,Borsanyi2,Aoki1,Aoki2}. 
As a consequence, the very definition and numerical determination of screening mass are obscured by the complications of the non-abelian nature of QCD and the strong coupling.
Nevertheless, let us note that parametrization of the real part of effective potential $U(r)$ can reproduce the lattice data quite well \cite{Dixit}. 
In this paper, to identify a running coupling constant and a Debye screening mass of quasiparticles, we are fitting the spatial pair distribution functions with a form, corresponding to a screened Coulomb potential.  
The pair distribution functions also allow us to identify the gluon bound states in the form 
of glueballs, but only at densities related to the average inter-particle distances more than $r_s\ge 0.4$~fm. ($r_s=\sqrt[3]{3/4\pi n}$, n is the density of all quasiparticles).  

For detailed studies of the quark, antiquark and gluon color screenings we have also developed the new spatial color pair correlation function with respect to the inter-particle distance and introduced related new color screening masses as function of temperature.
	
The quasiparticle model, considered here,  also allows us to investigate the kinetic properties of QGP.
For this purpose the quantum color dynamics in the color phase space has been developed in \cite{EbelForFil}, moreover  diffusion coefficient and shear viscosity of QGP, calculated there, are in a quite good agreement with available data. 
In the present paper in the framework of color phase space Wigner approach to quasiparticle model we propose Monte Carlo calculations of the quark and gluon momentum distribution functions for strongly coupled QGP in thermal equilibrium at zero baryon chemical potential.
To do this we rewrite the Wigner function of QGP in the form of color path integrals. 
To integrate over color variables, we develop a procedure of sampling the color quasiparticle variables in accordance with the Haar measure of the SU(3) group with the quadratic and cubic Casimir conditions.
The developed approach self-consistently
takes into account Fermi (Bose) statistics of quarks (gluons).
	
The paper is organized as follows. 
Section II deals with assumptions of the quasiparticle description of hot QCD.
In section III and IV we discuss the path integral representation of the matrix elements of the density operator and the Wigner functions for canonical ensemble correspondingly.
Section V is devoted to the brief description  of the Monte Carlo simulations.  
Results of simulations of the pair distribution functions are considered in section VI.
In section VII the spatial pair distribution functions, the Debye screening mass and the running coupling constant are discussed.
In sections VIII the new color pair correlation functions are presented and the related color screening mass is discussed.
Section IX deals with the quantum ``tails'' of QGP momentum distribution functions.

\section{Assumptions of the model}
The basic assumptions of the considered quasiparticle model are similar to those in \cite{shuryak1,shuryak2017,EbelForFil}.\\
1) Masses of quasiparticles $m$ are of order or higher than the mean kinetic energy per particle; this assumption is based on the analysis of QCD lattice data \cite{Lattice02,LiaoShuryak,Karsch:2009tp}. \\
2) We consider the model with quarks of three flavors ; for  simplicity, we assume the masses of $u$, $d$ and $s$  quarks to be equal; as for gluon quasi- particles, we allow their masses to differ from that of quarks (heavier). \\
3)The interparticle interaction is dominated by color-electric Coulomb potential; color-magnetic effects are neglected as subleading ones. \\
4) Since the color charges are large, the color operators are replaced by their average values, i.e. by Wong’s classical color vectors [eight-dimensional (8D) in SU(3)] with the quadratic and cubic Casimir conditions \cite{Wong}.

Applicability of this approach has been discussed in \cite{LM105,shuryak1} in details.
Our approach differs from \cite{LM105,shuryak1} by a quantum treatment to quasiparticles instead of the classical one.
This model requires the following quantities as functions of temperature $T$ and quark chemical potential $\mu_q$: \\
1) quasiparticle masses $m_q$ for quarks and $m_g$ for gluons; \\
2) the coupling constant $g^2$, or $\alpha_s = g^2/4\pi$. \\ 
It would be ideal if the input quantities are deduced from lattice QCD data or from other appropriate model. 
However, this task is still quite ambiguous presently, so in the present simulations we take 
only a possible set of parameters ($m_g$ and $m_q$ ) given by the HTL perturbative approach \cite{Lebel96}:
\begin{widetext}  
	\begin{eqnarray}
		\label{m_g-qg-T}
		m_g^2 (\{\mu_{q}\},T) =
		\frac{1}{12}\left((2N_c+N_f)T^2 +
		\frac{3}{\pi^2}\sum_{q=u,d,s}
		\mu_{q}^2\right)g^2(\{\mu_{q}\},T \gg T_C) \,
		\\
		\label{m_q-qg-T}
		m_{q}^2(\{\mu_{q}\},T) =
		\frac{N_g}{16N_c}\left(T^2+\frac{\mu_{q}^2}{\pi^2}\right)g^2(\{\mu_{q}\},T \gg T_C),
	\end{eqnarray}
\end{widetext} 
where $N_f$ is the number of quark flavors which can be excited, $N_c = 3$ for SU(3) group, and $g^2$ is square of 		the QCD running coupling constant, generally depending on $T$ and all $\mu_q$.
All masses depend on combinations of
$ z_{g}=\left(T^2 +
\frac{3}{\pi^2 (2N_c+N_f)}\sum_{q=u,d,s}
\mu_{q}^2\right)^{1/2} $
and $ z_{q}=\left(T^2+\frac{\mu_{q}^2}{\pi^2}\right)^{1/2} $
rather than on two independent variables $T$ and $\mu_{q}$.
It is also reasonable to assume that $g^2$ is a function of this single variable $z_{g}$ , because $g^2$ is related to the whole system rather than one specific quark flavor.
Then we can use ``one-loop analytic coupling constant'' \cite{Shirkov1,Shirkov2,Prosperi}:
\begin{eqnarray}
	\alpha_s (Q^2) = \frac{4\pi}{11-(2/3)N_f} \left[
	\frac{1}{\ln (Q^2/\Lambda_{\rm QCD}^2)} + \frac{\Lambda_{\rm QCD}^2}{\Lambda_{\rm QCD}^2-Q^2}\right],
	\nonumber
\end{eqnarray}
where $Q$ is the momentum transfer, $ \Lambda_{\rm QCD} = 206$ MeV is the QCD scale, $N_f = 3$ is the number of flavors and $Q$ is replaced by $2\pi z_g$.
	
We consider a multi-component QGP consisting of $\tilde{N}$ color quasiparticles: $N_g$ gluons, $N_q$ quarks
and $N_{\overline q}$ antiquarks. 
The Hamiltonian of this system is $\hat{H}=\hat{K}+\hat{U}^C$ with the kinetic and color Coulomb interaction parts:
\begin{eqnarray}
	\label{Coulomb}
&&		\hat{K}=\sum_i \sqrt{\hat{\bf p}^2_i+m^2_i(T,\mu_i)},
\nonumber\\&&
	\hat{U}^C=\frac{1}{2}\sum_{i\neq j}
	\frac{g^2(T,\mu_i)
	(Q_i \cdot Q_j)} 
	{4\pi| {\bf x}_i-{\bf x}_j|}.
\end{eqnarray}
Here $i$ and $j$ run over all quark and gluon quasiparticles,
$\mu_i$ are their chemical potentials,
$i,j=1,\ldots,\tilde{N}$, $\tilde{N}=N_q+N_{\overline q }+N_g$,
$N_q=N_u+N_d+N_s$ and $N_{\overline q }=N_{\underline u}+N_{\underline d}+N_{\underline s}$ are total numbers of quarks and antiquarks of all flavors ($u$, $d$, $s$),
3D vectors ${\bf x}_i$ are quasiparticle dimensionless spatial coordinates,  $g^2(T,\mu_i)/4\pi$ is  coupling constant, 
the $Q_i$ denotes the Wong's quasiparticle color variable (8D-vector in the group $SU(3)$),
$(Q_i \cdot Q_j)$ denotes the scalar product of color vectors.
Non-relativistic approximation for potential energy is used, while for kinetic energy we still keep the relativistic form,  since the temperature is not negligible in comparison with quasiparticle masses.
The equation of eigenvalues of this Hamiltonian is usually called  “spinless Salpeter equation”.	
The grand canonical ensemble with given temperature, net-quark-number ($\mu_q$ ), strange ($\mu_s$ ) chemical potentials, and fixed volume $V$ is completely described by grand partition function:
	%
	%
\begin{widetext}
	\begin{eqnarray}
		Z\left(\mu_q,\mu_s,\beta,V\right)=
		\sum_{\{N\}}
		\frac{\exp\{\mu_q(N_q-N_{\overline q })/T\}\;\exp\{\mu_s(N_s-N_{\underline s})/T\}}%
		{N_u!\;N_d!\;N_s!\;N_{\overline q }!\;N_{\underline d }!\;N_{\underline s }!\;N_g!}
		Z\left(\{N\},V,\beta\right), 
		\label{Gq-def}
	\end{eqnarray}
\end{widetext}
where $\{N\}=\{N_u,N_d,N_s,N_{\underline u },N_{\underline d }N_{\underline s } ,N_g\}$.
In equation (\ref{Gq-def}) we explicitly wrote sum over different quark flavors (u,d,s).
The sum over quark degrees of freedom is understood in the same way below.
Usual choice of the strange chemical potential is $\mu_s=-\mu_q$ (nonstrange matter),
such that the total factor in front of $(N_s-N_{\underline s })$ is zero,	$\beta=1/T$ is the reciprocal temperature.
Therefore, we omit $\mu_s$ from the list of variables below.
	
The partition function in canonical ensemble $Z\left(\{N\},V,\beta\right)$ and related thermodynamic properties of many particle system are defined by  diagonal matrix elements of the density operator ${\hat \rho} = \exp (-\beta{\hat H})$:
\begin{widetext}
	\begin{eqnarray}
		Z\left(\{N\},V,\beta\right) &=&
		\sum_{\sigma,\acute{\sigma}} \int \rm  dx\; \rm d\acute{x}\; \rm d\mu Q \; \rm d\mu \acute{Q} \;
		\delta_{\sigma,\acute{\sigma}}\delta(x-\acute{x})\delta(Q-\acute{Q})
		\langle x,Q,\sigma|e^{-\beta\hat H(Q)}|\acute{x},\acute{Q},\acute{\sigma}\rangle \,
		\nonumber\\
		&=&\sum_{\sigma} \int	\rm  dx\; \rm d\mu Q \;\rho(x,Q, \sigma),
		\label{Z-def}
	\end{eqnarray}
\end{widetext}
where $x$, $\sigma$, $Q$ denote the multi-dimensional vectors, related to spatial, spin and color degrees of freedom of $N$ quasiparticles with related flavor indexes respectively.
The summation over $\sigma$, spatial ($\rm dx \equiv \rm d^{3} x_1 \dots \rm d^{3} x_N $)
and color ($\rm d\mu Q\equiv \rm d\mu Q_1 \dots \rm d\mu Q_N $) integrations
run over all individual degrees of freedom of the quasiparticles,
while $\rm d\mu Q_i$ denotes integration over SU(3) group Haar measure \cite{LM105,QGP1}.

\section{Path integral representation of the density matrix}
The exact matrix elements of density operator $\rho=\rm e^{-\beta {\hat H}}$ of interacting quantum system can be constructed using a path integral approach \cite{Feynm,zamalin}, based on operator identity 
$\rm e^{-\beta \hat{H}}=\rm e^{-\varepsilon {\hat H}} \cdot \rm e^{-\varepsilon {\hat H}}\dots \rm e^{-\varepsilon {\hat H}}$, 
where the r.h.s. contains $M$ identical factors with $\varepsilon = \beta/M$, allowing us to rewrite the integral in equation (\ref{Z-def}) as follows:
\begin{widetext} 
	\begin{eqnarray}
	&&\sum_{\sigma } \int\limits \rm  dx \rm d\mu Q\,
		\rho(x,{\rm Q},\sigma) =
		\sum_{\sigma} \int\limits \rm  dx \rm d\mu Q \int\limits \rm  dx^{(1)}\rm d\mu Q^{(1)}\dots
		\rm dx^{(M-1)}\rm d\mu Q^{(M-1)} \, \rho^{(1)}\cdot\rho^{(2)} \, \dots \rho^{(M-1)} 
		\nonumber\\&& 
		\times \sum_{{\rm P}_q} \sum_{{\rm P}_{ \overline{q}}}\sum_{{\rm P}_g}(-1)^{\kappa_{{\rm P}_q}+ \kappa_{{\rm P}_{\overline{q}}}}
		\sum_{\sigma'} {\cal S}(\sigma, {\rm P}_{q\overline{q}g} \sigma^\prime)\delta_{\sigma',\sigma} \,
		\nonumber\\&&
		\times \int\limits {\rm d}x^{(M)} {\rm d}\mu {\rm Q}^{(M)} \delta(x-{\rm P}_{q\overline{q} g}x^{(M)})
		\delta({\rm Q}-{\rm P}_{q\overline{q}g}{\rm Q}^{(M)})  \rho^{(M)}, 
		\label{Grho-pimc0}
	\end{eqnarray}
\end{widetext} 	
where $x \equiv x^{(0)}$, ${\rm Q} \equiv {\rm Q}^{(0)}$, spin gives rise to the spin part of the density matrix
(${\cal S}$) with exchange effects accounted for by the permutation
operators ${\rm P}_q$, ${\rm P}_{ \overline{q}}$ and ${\rm P}_g$ acting on the quasiparticle indexes of quarks, 
antiquarks and gluons in $x^{(M)}$, ${\rm Q}^{(M)}$ and the spin projections $\sigma'$,  $ {\rm P}_{q\overline{q}g}={{\rm P}_q}{{\rm P}_{ \overline{q}}}{{\rm P}_g}$. 
	
The sum runs over all permutations with parity $\kappa_{P_q}$ and $\kappa_{P_{\overline{q}}}$, while
\begin{eqnarray}
&&\rho^{(m)} \equiv
		\rho\left(x^{(m-1)},{\rm Q}^{(m-1)};x^{(m)},{\rm Q}^{(m)};\{N\};\varepsilon \right) 
		\nonumber\\&&		
=\langle x^{(m-1)}|e^{-\varepsilon {\hat H}}|x^{(m)}\rangle\delta({\rm Q}^{(m-1)}-{\rm Q}^{(m)})
		\label{rho(m)}
\end{eqnarray}
is the off-diagonal element of the density matrix.
Since the color charge is treated classically, we keep only diagonal terms
($\delta({\rm Q}^{(m-1)}-{\rm Q}^{(m)})$) in color degrees of freedom.
Here each quasiparticle is presented by … set of coordinates 	$\{x_i^{(0)}, \dots, x_i^{(M-1)}\}$, called ``beads'', in units of $\lambda_a=\sqrt{2\pi\varepsilon /m_a}$, $a=q,\overline{q},g$, $\hbar=k_B=c=1$)	
and a 8-dimensional color vector $Q_i^{(0)}$ in the $SU(3)$ group.
Thus, all ``beads'' of each quasiparticle are characterized by the same spin projection, flavor and color charge.
Notice that masses and coupling constant $g^2(T,\mu_i)/4\pi$ in each $\rho^{(m)}$ are the same as those 
for the original quasiparticles, i.e. these are still defined by the actual temperature $T$. 
Details of analytical calculation of matrix elements $\rho^{(m)}$ are presented in \cite{EbelForFil,QGP1,QGP2}.	
	
The main advantage of this approach is that it allows us to use perturbation theory to obtain approximation for density matrices $\rho^{(m)}$, which is applicable due to smallness of artificially introduced factor $1/(M)$.
Each factor $\rho^{(m)}$ should be calculated with the accuracy of order of $1/M^{\theta}$ with $\theta > 1 $, because in this case the error of the whole product in the limit $M \to \infty$ tends to zero.
	
\section{The Wigner function for canonical ensemble} 
Now we are going to obtain a new path integral representation of Wigner functions in the color phase space, which allows us numerical simulations of strongly coupled quantum systems of particles in canonical ensemble \cite{Feynm,LarkinFilinovCPP,JAMP,Wiener,NormanZamalin,zamalin}. 	
The Wigner function of many-particle system in canonical ensemble can be defined as a Fourier transform \cite{Wgnr,Tatr} of the matrix element of the density operator \cite{QGP1,QGP2} in the coordinate representation:
\begin{widetext} 
\begin{eqnarray}\label{pathint_wignerfunctionint2}
&&W(p,x,{\rm Q})=\sum_{\sigma,\acute{\sigma}} 
\int \rm  d\xi\;  \rm d\mu \acute{Q} \; \;
\delta_{\sigma,\acute{\sigma}}\delta(Q-\acute{Q})\exp (i\langle \xi |p\rangle)
\langle x+\xi/2,Q,\sigma|e^{-\beta\hat H(Q)}|x-\xi/2,\acute{Q},\acute{\sigma}\rangle \,
\nonumber \\&&
=\frac{C(M)}{Z\left(\{N\},V,\beta\right)}
\sum_{\sigma}\sum_{{ {\rm P}_q}{{\rm P}_{ \overline{q}}}{{\rm P}_g}} (-1)^{\kappa_{{\rm P}_q}+ \kappa_{{\rm P}_{ \overline{q}}}}
{\cal S}(\sigma, {{\rm P}_{q\overline{q}g}} \sigma^\prime)\big|_{\sigma'=\sigma}\,
\int\limits {\rm d}\mu {\rm Q}^{(M)} \delta({\rm Q}-{\rm P}_{q\overline{q}g}Q^{(M)})
\nonumber \\&&
\times \int {\rm d} \xi \,
\int {\rm d}q^{(1)} \dots {\rm d}q^{(M-1)}\, {\rm d}\mu {\rm Q}
\exp\Biggl\{-\pi \frac{\langle \xi|{\rm P}_{q\overline{q}g}+E |\xi \rangle}
{2M } + i\langle \xi |p\rangle
-\pi \frac{|{\rm P}_{q\overline{q}g} x-x|^2}{M}
\nonumber\\&&
-\sum\limits_{m = 0}^{M-1}
\biggl[\pi | q^{(m)}-q^{(m+1)}|^2 +
\varepsilon U\biggl(({\rm P}_{q\overline{q}g} x-x)\frac{m}{M}+x + q^{(m)}
-\frac{ (M-m) \xi}{2M}+\frac{ m {\rm P}_{q\overline{q}g}\xi}{2M} \biggr)
\biggr]
\Biggr\} , 
\nonumber \\
\end{eqnarray}
\end{widetext} 
where $C(M)=M^{6\tilde{N}(M-1)/2}$ is constant and $q^{(M)}=q^{(0)}$. 
Details of analytical calculation of the matrix elements of density operator are presented in \cite{EbelForFil,QGP1,QGP2}.
Here the interaction energy $U$ is the sum of the 
two-particle color quantum Kelbg potentials. 
The antisymmetrization for quarks and symmetrization for gluons takes into account quantum statistics.
Here we have replaced variables of integration $x^{(m)}$
for any given permutation ${{\rm P}_q}{{\rm P}_{ \overline{q}}}{{\rm P}_g}$ by relation
\begin{widetext} 
\begin{eqnarray}
x^{(m)} = ({\rm P}_{q \overline{q} g }x-x)\frac{m}{M}+x + q^{(m)}
-\frac{ (M-m) \xi}{2M}+\frac{ m {\rm P}_{q\overline{q}g}\xi}{2M} \,.
\label{pathint_variableschange}
\end{eqnarray}
\end{widetext} 
In equation~(\ref{pathint_wignerfunctionint2}) $E$ is the unit matrix, while the matrix presenting permutation
${{\rm P}_q}{{\rm P}_{ \overline{q}}}{{\rm P}_g}$ is equal to unit 
matrix with appropriately transposed columns.  
In equation (\ref{pathint_wignerfunctionint2}) $E$ is the unit matrix, while the permutation matrix ${{\rm P}_q}{{\rm P}_{ \overline{q}}}{{\rm P}_g}$  is obtained from $E$ by appropriate transposition of columns.

To avoid the problems with definition of the relativistic Wigner function, discussed in \cite{reveiw,Zav1,Zav2,Lar1}, we use here the non-relativistic limit for kinetic energy operator (\ref{Coulomb}).	
	
The expression for the Wigner function (\ref{pathint_wignerfunctionint2}) is inconvenient for Monte Carlo simulations since it does not contain explicit result of integration over $\xi$, even for free particles with $U(x) \equiv 0$.
In general, this integral can not be calculated
analytically. 
Exceptions are linear and quadratic functions $U (x)$, known as the linear and harmonic potentials correspondingly.	
To perform the integration over $\xi$ analytically and obtain an explicit expression for Wigner function , let us take the approximation for potential $U(x)$, given by the Taylor expansion up to the first or second order in $\xi$, strongly restricted by exponentially decaying factor with quadratic form of  $\xi$ in (\ref{pathint_wignerfunctionint2}).
This approximations were tested by calculations of thermodynamic values and ground state wave functions for quantum particle in 1D and 3D potential field \cite{LarkinFilinovCPP, JAMP}; they gives practically exact results even for potentials, which differ from linear or harmonic ones significantly.
		
As it was shown in \cite{JPA2017} for electromagnetic plasma, the main contribution to Fermi repulsion in (\ref{pathint_wignerfunctionint2}) comes from the pair permutations at moderate plasma degeneracy, when temperature is about Fermi energy.
This is the physical reason to take into account only pair permutations and neglect the others.
In this approximation the Wigner function can be presented in the next form:
\begin{widetext} 	
	\begin{eqnarray}
		\label{pathint_wignerfunctionint4}
	&&W(p,x,{\rm Q}) \approx\,
		\frac{C(M)}{Z\left(\{N\},V,\beta\right)}
		\int \rm dq^{(1)} \dots \rm dq^{(M-1)}\,
		\nonumber\\&&
		\times	\exp\Biggl\{
		-\sum\limits_{m = 0}^{M-1}
		\biggl[\pi | q^{(m)}-q^{(m+1)}|^2 +
		\varepsilon U\biggl(x + q^{(m)}
		) \biggr] \Biggr\}
\exp\Biggl\{\frac{M}{4 \pi}
		\Biggr|i p + \frac{\varepsilon}{2}\sum\limits_{m = 0}^{M-1} \frac{ (M-2m) }{M}
		\frac{\partial U(x+q^{(m)})}{\partial x}
		\Biggr|^2\Biggr\}
		\nonumber\\&&
		\times\sum_{\sigma_q}\Biggl\{1-\sum_{l<t}^{N_q}\delta_{\sigma_{l,q}\sigma_{t,q}}
		\delta_{f_{l,q}f_{t,q}} \delta(({\rm Q}_{l,q}-{\rm Q}_{t,q}))
		\exp(-2\pi\frac{|x_{l,q}-x_{t,q}|^2}{M})
		\delta \biggl(\frac{(\tilde{p}_{l,q}-\tilde{p}_{t,q})\sqrt{M}}{2\pi}
		\biggr)\Biggr\}
		\nonumber\\&&
		\times\sum_{\sigma_{\overline{q}}}\Biggl\{1-\sum_{l<t}^{\bar{N}_q}\delta_{\sigma_{l,{\overline{q}}}\sigma_{t,{\overline{q}}}}
		\delta_{f_{l,\overline{q}}f_{t,\overline{q}}} \delta(({\rm Q}_{l,\overline{q}}-{\rm Q}_{t,\overline{q}}))
		\exp(-2\pi\frac{|x_{l,{\overline{q}}}-x_{t,{\overline{q}}}|^2}{M})
		\delta \biggl(\frac{(\tilde{p}_{l,{\overline{q}}}-\tilde{p}_{t,{\overline{q}}})\sqrt{M}}{2\pi}
		\biggr)\Biggr\}
		\nonumber\\&&
		\times\sum_{\sigma_g}\Biggl\{1+\sum_{l<t}^{N_g}
		\delta(({\rm Q}_{l,g}-{\rm Q}_{t,g}))		
		\exp(-2\pi\frac{|x_{l,g}-x_{t,g}|^2}{M})
		\delta \biggl(\frac{(\tilde{p}_{l,g}-\tilde{p}_{t,g})\sqrt{M}}{2\pi}
		\biggr)\Biggr\},
	\end{eqnarray}
\end{widetext} 	
	where
	\begin{eqnarray}
		\tilde{p}_{t,a}=p_{t,a} + \frac{\varepsilon }{2 }\sum\limits_{m = 0}^{M-1}
		\frac{\partial U(x+q^{(m)})}{\partial x_{t,a}}.
		\nonumber
	\end{eqnarray}
Here $\delta_{f_i,f_j}$ are the Kronecker symbols, depending on flavor indexes $f_i$ of quasiparticles and taking values $u$, $d$ and $s$. 
For simplicity, we write here the expression, corresponding only to the linear terms in the Taylor expansion of color potentials.
	
To avoid difficulties, arising at Monte Carlo simulations due to presence of the delta-function in expression (\ref{pathint_wignerfunctionint4}), and to regularize integration over momenta, the positive Husimi distributions, being a coarse-grained Wigner function, can be used  with a Gaussian smoothing for small phase space cells of parameters $\Delta^2_x$ and $\Delta^2_p$  \cite{Tatr}.
	The final expression for Wigner function can be written in the form:
\begin{widetext} 	
	\begin{eqnarray}
		\label{pathint_wignerfunctionint5}
	&&W^H(p,x,{\rm Q}) \approx\,
		\frac{C(M)}{Z\left(\{N\},V,\beta\right) }
		\int \rm dq^{(1)} \dots \rm dq^{(M-1)}\,
\exp\Biggl\{
		-\sum\limits_{m = 0}^{M-1}
		\biggl[\pi | q^{(m)}-q^{(m+1)}|^2 +
		\varepsilon U(x + q^{(m)}
		) \biggr] \Biggr\}
		\nonumber\\&&
		\times \exp\Biggl\{\frac{M}{4 \pi}
		\Biggr| i p + \frac{\varepsilon}{2}\sum\limits_{m = 0}^{M-1} \frac{ (M-2m) }{M}
		\frac{\partial U(x+q^{(m)})}{\partial x}
		\Biggr|^2\Biggr\}
\sum_{\sigma}\exp (-\beta \sum_{l<t}^{N_q} v^{q}_{ lt})\exp (-\beta \sum_{l<t}^{\bar{N}_q}
		v^{{ \overline{q}}}_{ lt})\exp (-\beta \sum_{l<t}^{N_g}v^{{g}}_{ lt}),
		\nonumber\\
	\end{eqnarray}
\end{widetext} 	
where the final expression for the phase space pair pseudopotentials, accounting for quantum statistical effects, looks like:
\begin{widetext} 	
	\begin{eqnarray}
	&&v^{a}_{ lt}=-kT \ln\Biggl\{
		1-\delta_{\sigma_{l,a}\sigma_{t,a}}\delta_{f_{l,a}f_{t,a}}\exp \biggl(-\frac{|{\rm Q}_{l,a}-{\rm Q}_{t,a})|^2}{2\tilde{\Delta}^2_Q}\biggr)
		\nonumber\\&&
		\times\exp \biggl(-\frac{2\pi|x_{l,a}-x_{t,a}|^2 (1-\frac{\tilde{\Delta}^2_{a,x}/\lambda^2_a}{1+\tilde{\Delta}^2_{a,x}/\lambda^2_a})}{\lambda^2_a}\biggr)
		\exp \biggl(-\frac{|(\tilde{p}_{l,a}-\tilde{p}_{t,a})|^2\lambda^2_a}{(2\pi\hbar)^2(\tilde{\Delta}^2_p/\lambda^2_a)}\biggr)
		\Biggr\}, 
		\nonumber\\&&
		v^{g}_{ lt}=-kT \ln\Biggl\{
		1 + \exp \biggl(-\frac{ |{\rm Q}_{l,g}-{\rm Q}_{t,g})|^2}{2 \tilde{\Delta}^2_Q}\biggr)
		\nonumber\\&&
		\times\exp \biggl(-\frac{2\pi|x_{l,g}-x_{t,g}|^2 (1-\frac{\tilde{\Delta}^2_{g,x}/\lambda^2_g}{1+\tilde{\Delta}^2_{g,x}/\lambda^2_g})}
		{\lambda^2_g}\biggr) \exp \biggl(-
		\frac{|(\tilde{p}_{l,g}-\tilde{g}_{t,g})|^2\lambda^2_g}{(2\pi\hbar)^2\tilde{(\Delta}^2_p/\lambda^2_g)}\biggr) 
		\Biggr\}.
	\end{eqnarray}
\end{widetext} 	
Here $a=q,\overline{q}$, $\Delta^2_{a,x}=\frac{2}{(\pi^2-2)}$ and $z=1/\sqrt{2}$. 
To extent the region of applicability of obtained phase space pair pseudopotential, $\tilde{\Delta}^2_p$ and $\tilde{\Delta}^2_Q $ can be considered as fit functions with values much smaller than unity.
Our test calculations \cite{JPA2017} have shown that the best fit for $\tilde{\Delta}^2_p$ can be written in the form
$\tilde{\Delta}^2_p/\lambda^2_a=0.00505+0.056n\lambda_a^3$, while $\tilde{\Delta}^2_{a,x}$ and $\tilde{\Delta}^2_Q$ were of order $0.1$.

The pseudopotentials $v^{q}_{ lt}$ in the phase space allow us to avoid the famous 'fermionic sign problem' and to realize the Pauli blocking for quarks/antiquarks with the same spin, flavor and color.
The pseudopotentials $v^{g}_{ lt}$  provides Bose statistics for gluons.
Note also that the expression (\ref{pathint_wignerfunctionint5}) explicitly contains the term, related to the classical Maxwell distribution,  modified by terms accounting for influence of interaction on the momentum distribution function.
	
An average value of arbitrary quantum operator $\hat A$ can be written as Weyl symbol $A(p,x,Q)$, averaged over the color phase space with the 	Wigner function $W(p,x,Q)$ or $W^H(p,x,Q)$:
\begin{eqnarray}\label{wigfunc_average0}
\langle \hat A\rangle = \int{\rm d}\mu {\rm Q} \;\frac{{\rm d}p {\rm d}x } {(2\pi)^{6\tilde{N}}}
A(p,x,{\rm Q}) W^H(p,x,\rm Q),
\end{eqnarray}
where the Weyl symbol of operator $\hat A$ is:
\begin{eqnarray}\label{wigfunc_weylsymbol}
A(p,x,{\rm Q}) = \int \rm d\xi \rm e^{-\rm i\langle
	\xi|p\rangle /\hbar} \langle x-\xi/2|\hat A(Q)|x + \xi/2\rangle.
\end{eqnarray}
Weyl symbols of common operators like $\hat p$, $\hat x$, $\hat p^2$, $\hat x^2$, $\hat H$, $\hat H^2$ etc. can be easily calculated directly from the definition (\ref{wigfunc_weylsymbol}).
	
\section{Monte Carlo simulations of QGP distribution functions}
\label{s:model}
The basic idea of the Monte Carlo simulations of QGP is to construct a Markovian chain of different quasiparticle states in the color configuration space or in the color phase space \cite{EbelForFil}.
The computational procedure comprises two stages.
At the first stage a dominant, i.e., maximal, $\{N\}$-term in the sum of Eq.~(\ref{Gq-def})
is determined by calculations in the grand canonical ensemble.  
This term is indeed the dominant one in thermodynamic limit of the box volume ($V \to \infty$).
In grand canonical ensemble, the quasiparticle numbers in the simulation box are varied, i.e., the sequential states of the Markovian chain can differ from each other by numbers of quarks, antiquarks or gluons. 
Transitions between these states are the first type of Markovian elementary step. 
In the second type of elementary step, the coordinates of randomly chosen quasiparticle are changed.
The color variables are changed according to the SU(3) group Haar measure \cite{LM105,LM106,LM107,LM108,QGP1,EbelForFil} in the third type of the Markovian elementary step.
For simplicity, numbers of spins "up' and "down" are fixed and equal to each other for quarks (antiquarks). 
The Markovian chain is generated until complete convergence of calculated values is achieved.
This allows one to determine the average numbers of quarks, antiquarks, and gluons in the box volume at fixed temperature. 
Here, only the densities of each type of particles, i.e., average number of particles to box volume ratio, have physical sense.
Usually, after several million elementary steps, the average numbers of quasiparticles of each type become stable, and, for example, at zero baryon chemical potential, the average number of quarks is practically equal to the average number of antiquarks.
This equality can be considered as an inherent test of self-consistency of the calculations.
In this way, the dependence of quasiparticle density on baryon chemical potential is calculated in grand canonical ensemble according to (\ref{Gq-def}).

At the second stage the fixed numbers of quarks, antiquarks, and gluons have to be chosen  equal to the obtained average values at the first stage, and calculations were carried out in the canonical ensemble.
Wherein the second and the third types of the elementary Markovian step described above are used.
To generate the Markovian chain for integration in the color phase space (see (\ref{wigfunc_average0})), it is necessary to sample the momentum of some quasiparticle as the new Markovian elementary step \cite{JPA2017}.
\begin{figure}[htb]
  \begin{minipage}[ht]{1.0\linewidth}
  		\includegraphics[width=7.5cm,clip=true]{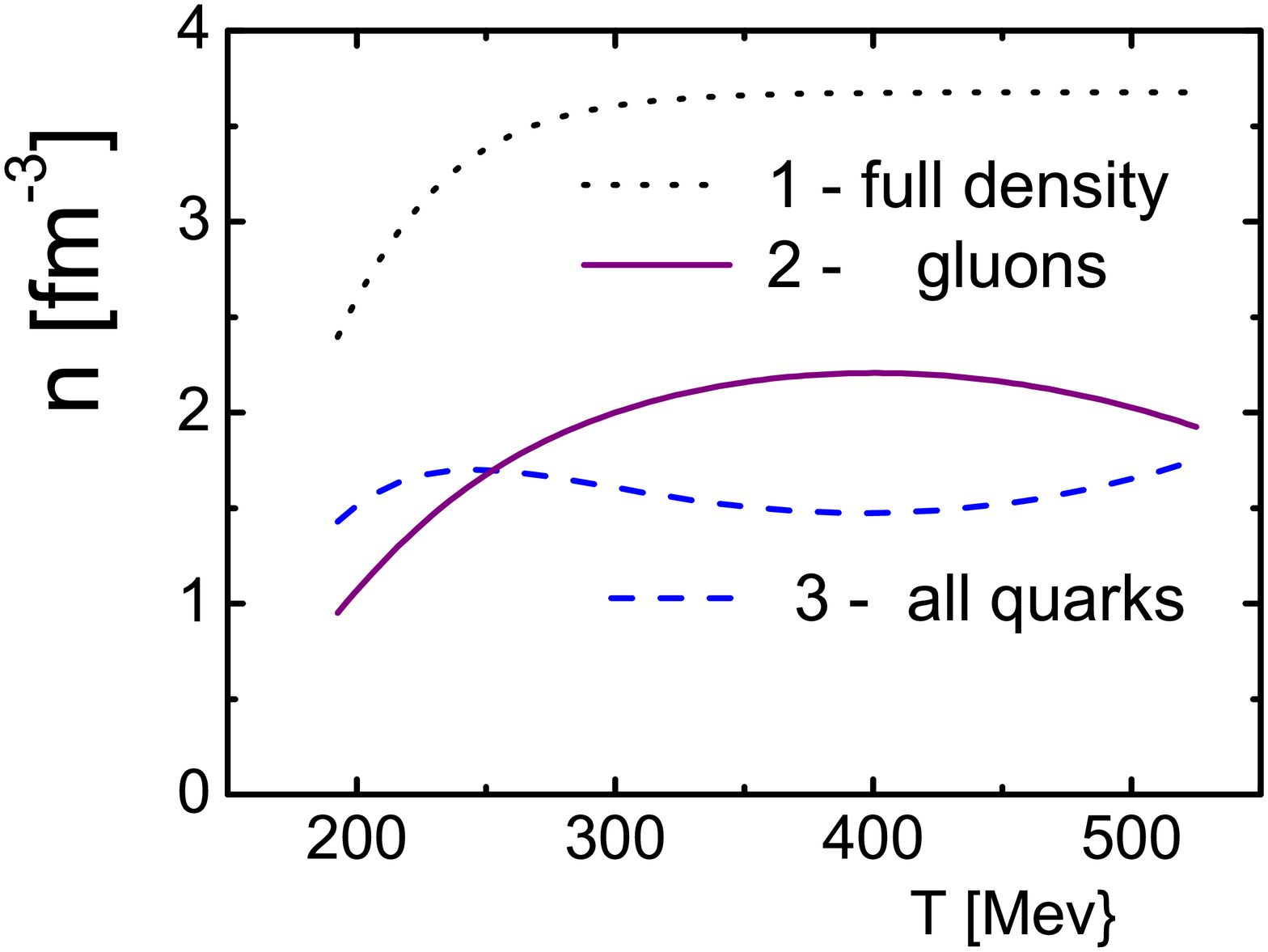}  \\(a)
  \end{minipage} \\
 \begin{minipage}[ht]{1.0\linewidth}
	\includegraphics[width=7.5cm,clip=true]{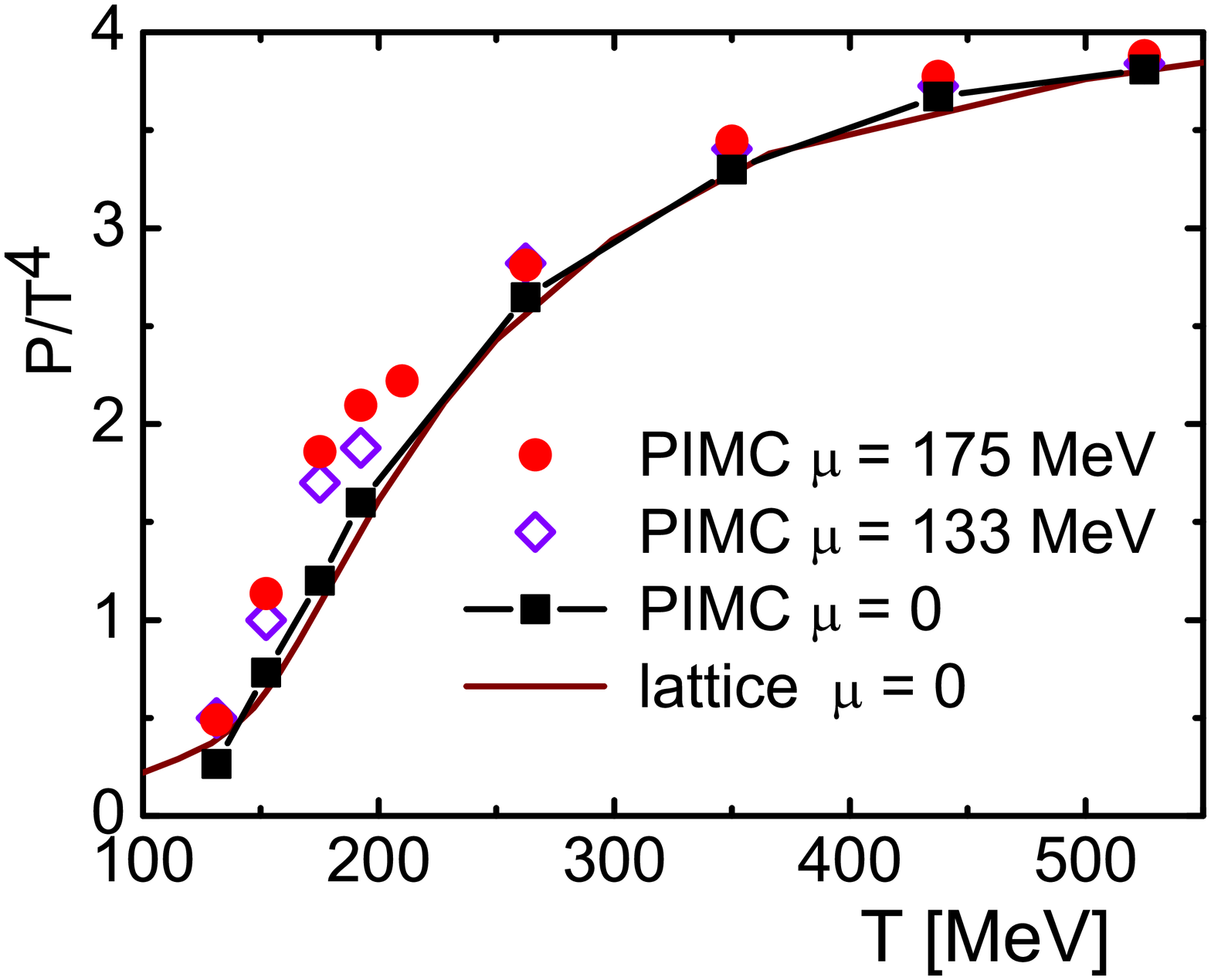} \\(b)
 \end{minipage}	\\
 \begin{minipage}[ht]{1.0\linewidth} 
   \includegraphics[width=7.5cm,clip=true]{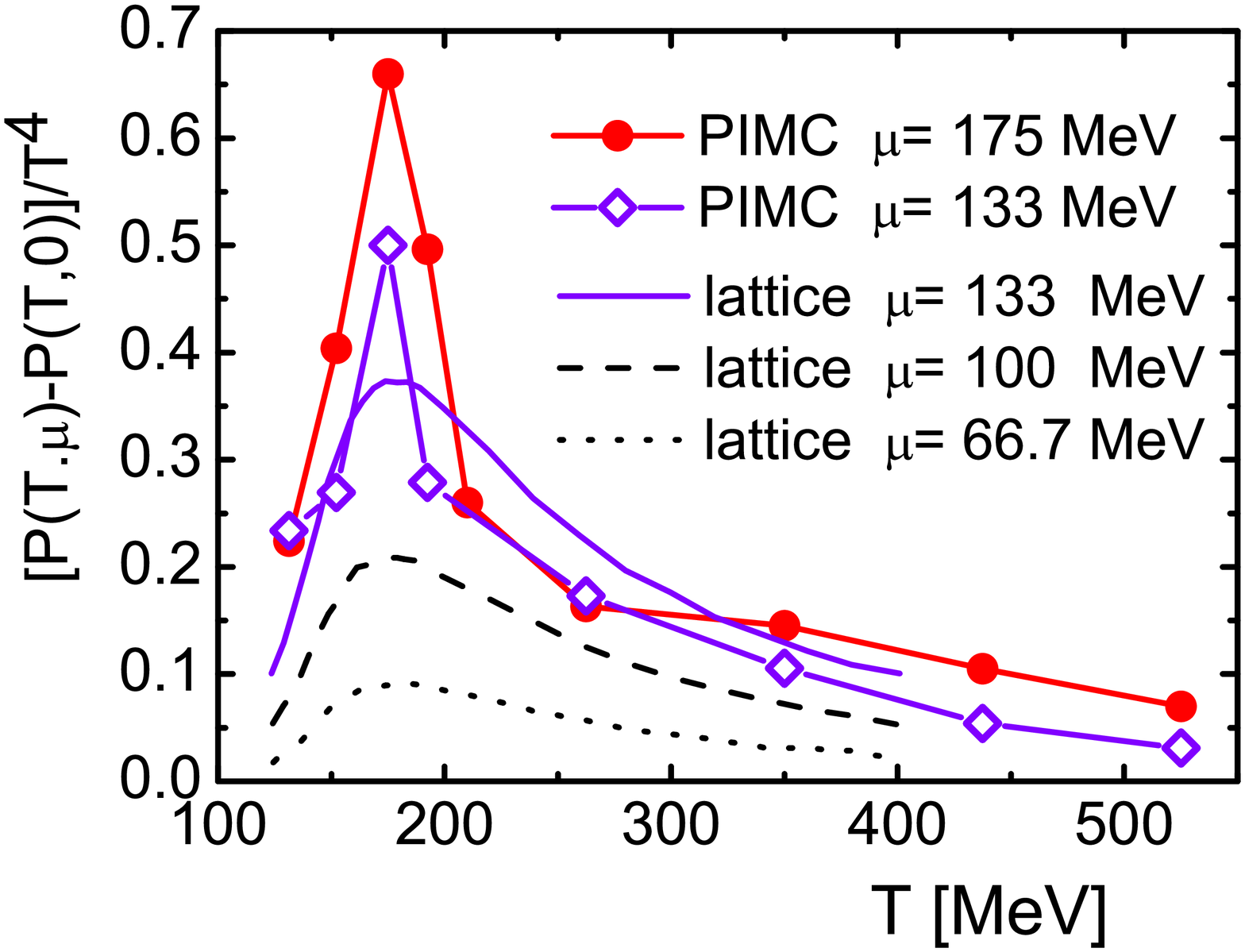} \\(c)
 \end{minipage}	   
	\caption{(Color online) (Plot (a)) The quark and gluon densities versus temperature at baryon chemical potential equal to zero. 
(Plot (b)) Equation of state of QGP for different baryon chemical potentials. 
(Plot (c)) Pressure for $\mu >0 $ in exceess of the pressure at $\mu =0$. 
The Monte Carlo (PIMC) results (symbols) are compared  to lattice data (lines) of Refs.~\cite{Borsanyi1,Borsanyi2}. 
}
	\label{density}
\end{figure}	

For reader convenience, Fig.~\ref{density} shows the results of the Monte Carlo calculations (PIMC) for averaged over spin and flavor variables, obtained in \cite{QGP1,QGP2} within quasiparticle approach, discussed above.
The top plot shows dependences of quark and gluon densities on temperature at zero baryon chemical potential ($ n_q=n_{\overline{q}}$).
For different chemical potentials the central and bottom plots present the equations of state and the pressure difference compared to the limit of zero chemical potential.
As  it follows from Fig.~\ref{density}, for chemical potentials $\mu/T_c=133/175<1$ our approach agrees quite
well with lattice QCD data based on a Taylor expansion around $\mu=0$.

In this paper we calculate the spatial and momentum distribution functions of QGP quasiparticles in canonical ensemble, using the quark, antiquark and gluon densities discuissed above.

\section{Pair distribution functions}
To undestand physical properties of QGP, let us start from consideration of spatial arrangement of  quasiparticles by discussing the pair distribution functions (PDF) $\rm g_{ab}(r)$, which are obtained by integration of the Wigner functions over all variables except coordinates of two quasiparticles of the types ’a’ and ’b’:
\begin{eqnarray}\label{g-def}
&&\rm g_{ab}(|{\bf R}_1-{\bf R}_2|) \frac{N_aN_b}{V^2} = 
\frac{1}{Z} 
\sum_{\sigma}
\sum_{i,j,i\neq j}
\delta_{a_i,a}\, \delta_{b_j,b}  \int \rm dp
\nonumber\\&&
\times  \rm dr d\mu  Q \delta({\bf R}_1-{\bf r}_i)\,\delta({\bf R}_2-{\bf r}_j)\;W^H(p,x,Q),  
\end{eqnarray}
where $a(a_i)$ and $b(b_j)$ can take values $q,\overline{q}$ or $g$.	
Functions $\rm g_{ab}(r)$ give probability density to find a pair of quasiparticles of types a and b at the distance $r$ between them as the PDFs depending only on the difference between coordinates due to translational invariance of the system.
\begin{figure}[htb]
\begin{minipage}[ht]{1.0\linewidth}	
	\includegraphics[width=6.5cm,clip=true]{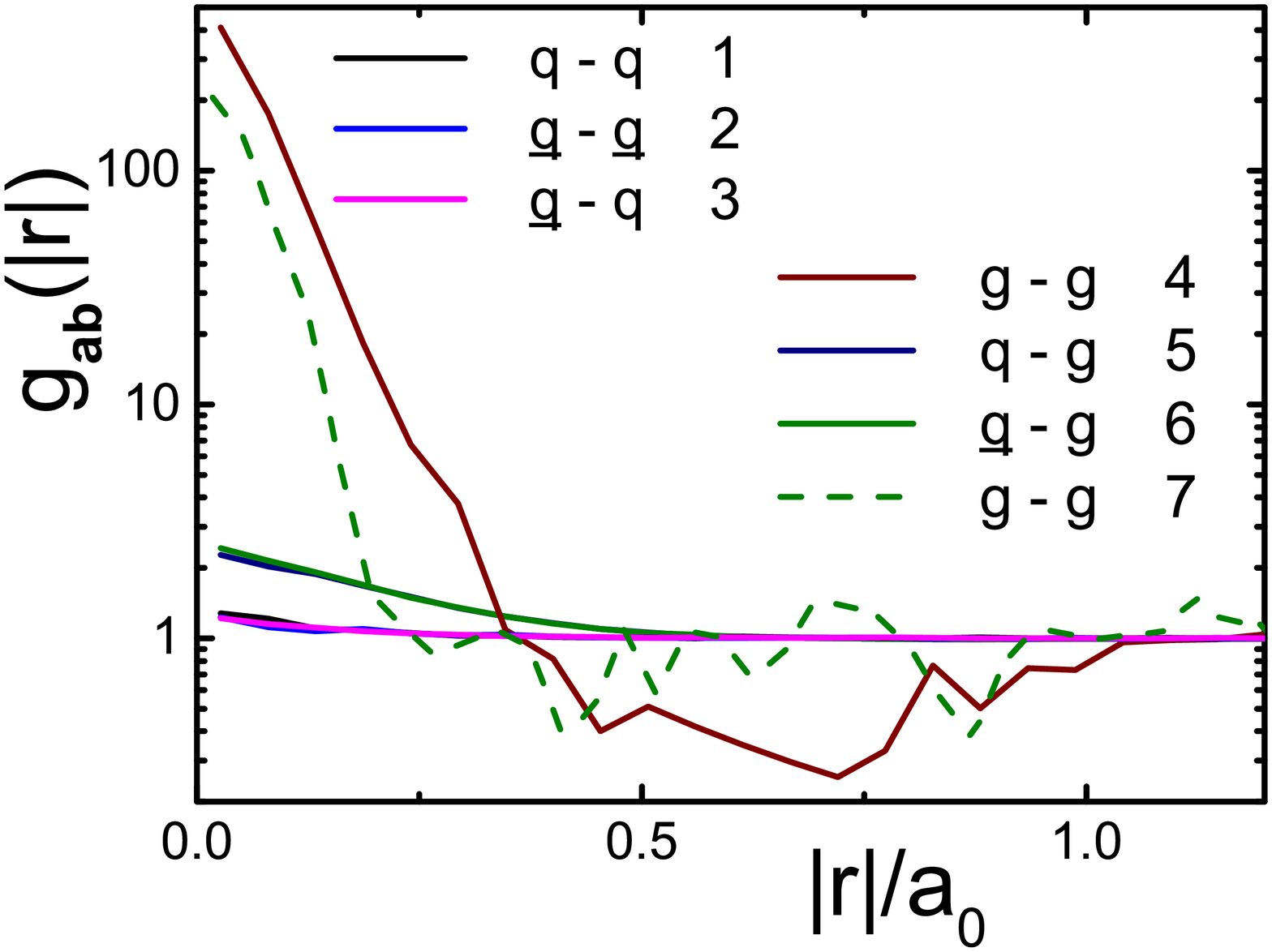}  \\(a)
\end{minipage} \\
\begin{minipage}[ht]{1.0\linewidth}	
	\includegraphics[width=6.5cm,clip=true]{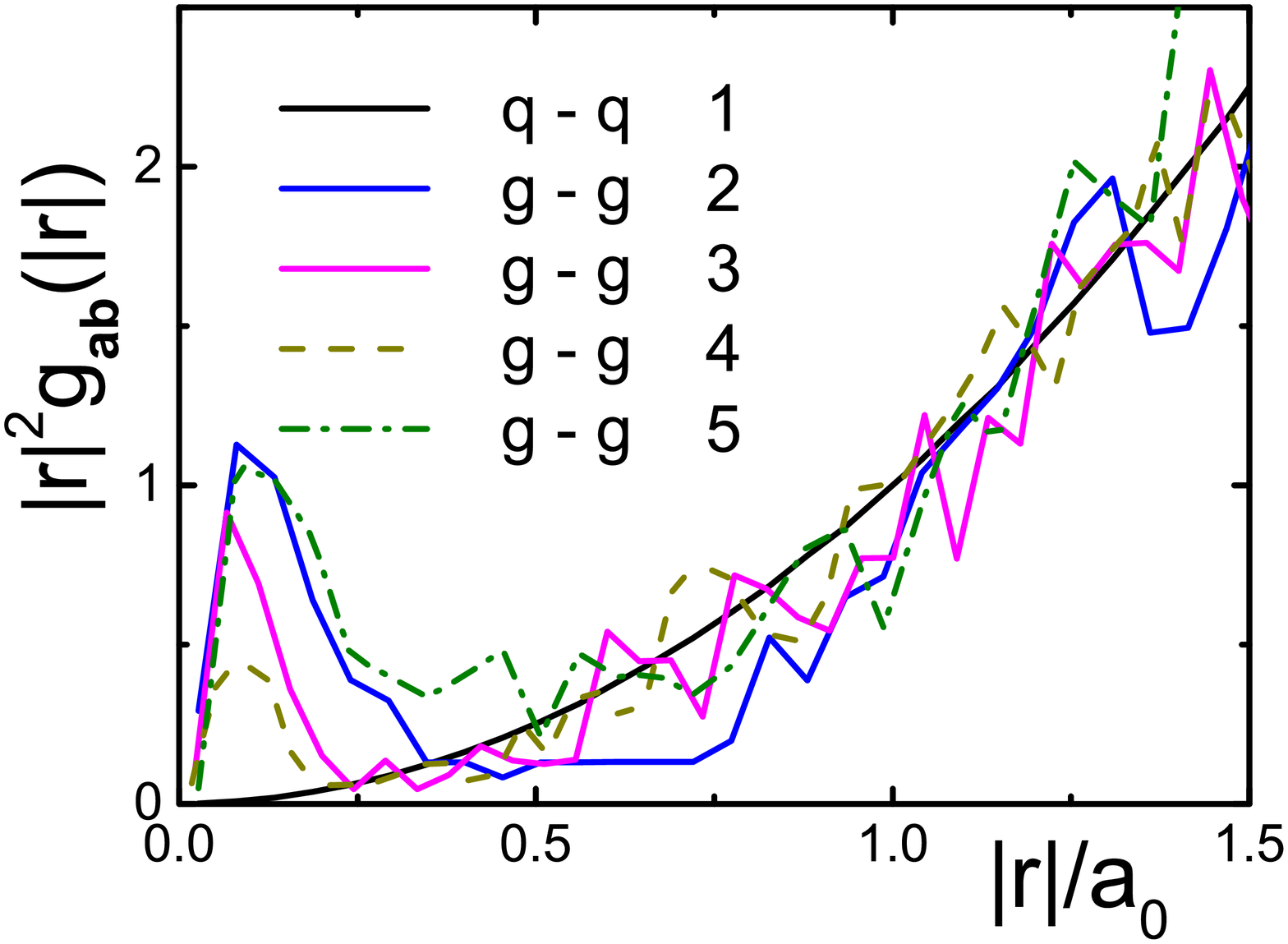}  \\(b)
\end{minipage} \\
	\caption{(Color online) (Plot (a)) The pair distribution functions $\rm g_{ab}(r)$ of the quark and gluon quasiparticles at temperatures $T=4T_c$ at baryon chemical potential equal to zero (a,b=quark, antiquark or gluon, $a_0=1.16$~fm). Lines: 1-6 ~-~ $r_s=0.6$~fm; 7 - $r_s=0.4$~fm. 
	(Plot (b))  Product $r^2 \rm g_{ab}(r)$. Lines for $T=4T_c$: 
	1,2 - $r_s=0.6$~fm; 3 - $r_s=0.5$~fm; 4 - $r_s=0.4$~fm.  
	Line 5 - $T=2T_c$, $r_s=0.6$~fm. 
	}
	\label{crt4}
\end{figure} 
In a non-interacting classical system $\rm g_{ab}(r)\equiv 1$, whereas interactions and  quantum effects result in 
a re-distribution of the particles.
Results for PDFs at temperature $T/T_c=2$ and  $T/T_c=4$ and different average interparticle distances $r_s$
($r_s=\sqrt[3]{3/(4\pi n)}$, n is the density of all quasiparticles)  
are shown on Fig.~\ref{crt4}. 
The PDFs of identical particles are presented by lines 1, 2, 4 and 7 on the top plot of Fig.~\ref{crt4}, while other lines show PDFs of the quasiparticles of different types.
All PDFs reveal a similar behavior. 
At the distances $r/a_0 \ge 0.5$ all PDFs tends to unity (i.e. the ideal gas limit), while near zero
PDFs are monotonously growing.
This behavior of the PDFs at small distances is manifestation of effective pair attraction of quarks, antiquarks and gluons. 
This attraction suggests that the color vectors of nearest neighbor quasiparticles of any type are anti-parallel. 
QGP lowers its total energy by minimizing the color Coulomb interaction energy via a spontaneous ``anti-ferromagnetic or colorless'' ordering of color vectors. 
This may result in clustering of quarks, antiquarks and gluons and, possible, appearance of  bound states.
Such low-distance behavior is also observed in a non- ideal dense astrophysical electron-ion plasma and in a nonideal electron-hole plasmas in semiconductors \cite{EbelForFil}. 

We can also notice fundamental difference between quark and gluon PDFs (antiquark PDFs are identical to the quark PDFs). 
The short-distance attraction is stronger for gluon-gluon and gluon-(anti)quark pairs	than for (anti)quark-(anti)quark ones because of the corresponding difference in values of quadratic Casimir invariants $\breve{q}_2$ \cite{EbelForFil}, 
which determine the maximal values of the effective color charge scalar products
$(\rm Q_i \cdot \rm Q_j)$ in color Kelbg (Coulomb) potentials:
for gluon-gluon pairs $\left|(\rm Q_g \cdot \rm Q_g)\right|_{max} = 24$,
for gluon-(anti)quark pairs $\left|(\rm Q_g \cdot \rm Q_q) \right|_{max}=
\left| (\rm Q_g \cdot \rm Q_{\overline{q}} ) \right|_{max}\approx$ 10,
and for (anti)quark-(anti)quark pairs
$\left| (\rm Q_q \cdot \rm Q_q ) \right|_{max}=
\left| (\rm Q_{\overline{q}} \cdot \rm Q_{\overline{q}} ) \right|_{max}=\left| ( \rm Q_q \cdot \rm Q_{\overline{q}} ) 			\right|_{max}=$ 4. 

The short-distance correlation implies formation of the gluon-gluon 
clusters, which are uniformly distributed in space (see ${\rm g}_{gg}(r)$ at large distance). 
In case of the gluon-gluon clusters we can even talk about $gg$-bound states (i.e. glueballs) due to the following well known statements of quantum mechanics for two particles.
The gluon-gluon PDFs can be formed either by correlated scattering states or by bound states of quasiparticles, depending on the relative fractions of these states.
However, strictly speaking, there is no clear subdivision into bound and free “components” in plasma media due to mutual overlap of the quasiparticle clouds.
In addition, there is not any rigorous criterion for a bound state at high densities due to the strong affection of the surrounding plasma.

Nevertheless, a rough estimate of existance and even fraction of quasiparticle bound states  \cite{ElHol}  can 	be obtained	by  the following reasonings. 
The product $r^2 \rm g_{ab}(r)$ has sense of probability to find a pair of quasiparticles at the distance $r$ between them.
On the other hand, the corresponding quantum mechanical probability is the product of $r^2$ and two-particle Slater sum
\begin{equation}
	\label{slsm}
	\Sigma_{ab}=8\pi
	^{3/2}\lambda_{ab}^{3}\sum_{\alpha}|\Psi_{\alpha}(r)|^{2}\exp(-\beta
	E_{\alpha})
	= \Sigma_{ab}^{d}+\Sigma_{ab}^{c},
\end{equation}
where $E_{\alpha}$ and $\Psi_{\alpha}(r)$ are the energy (without center of mass energy) and the wave function of a quasiparticle pair correspondingly.
$\Sigma_{ab}$ is, in essence, the diagonal part of the corresponding density matrix.
In Eq.~(\ref{slsm}) the summation runs over all possible states $\alpha$ 
with contributions from the discrete ($\Sigma_{ab}^{d}$) and continuous ($\Sigma_{ab}^{c}$) parts of the spectrum.

At the temperatures smaller than the binding energy and at distances smaller than or of the order of several bound state radii the main contribution to the Slater sum comes from bound states and the product $r^2\Sigma_{ab}^d$ is sharply peaked at distances around the Bohr radius \cite{ElHol}.   
Similarly, in QGP the product $r^2\rm g_{gg}(r)$ forms distinct maximum, which can be interpreted as  evidence of bound states of $gg$ pairs or clusters.
Sharp peak on the right plot of Fig.~\ref{crt4} at small distances indicates existence of the bound states; it demonstrates the fast decreasing fraction of the glueballs with increasing density at temperatures $T=4T_c$.
One can notice that at density related to $r_s=0.6$~fm fraction of bound states almost is not changed while temperature decreases from $T=4T_c$ to $T=2T_c$.

\section{Debye screening mass and running coupling constant} 
When studying the system by numerical methods, it is interesting to know how the free energy  depends on the interparticle distance.
The free energy surface along the selected coordinate is called the potential of mean force (PMF) \cite{Reith,Kirkwood1,Kirkwood2}.  
PMF can be obtained through Monte Carlo or Molecular Dynamics simulations, which examine how the energy of the system varies with a specific parameter.
For example, it can examine how the energy of the system changes as function of the distance between two given particles.   
This energy change $w^{(2)}(r)$ is an average work required to bring two particles from infinite distance to the distance $r$. 
The potential of mean force $w^{(2)}(r)$ is usually applied in the Boltzmann inversion method as a first guess for 		the effective pair interaction potential that ought to reproduce the correct pair distribution function \cite{Reith}.  	For low density of particles  the virial expansion in terms of bare potential $U_{ab}(r,T)$ gives  $w^{(2)}(r)=U_{ab}		(r,T)= -T \ln \rm g_{ab} (r,T)$ \cite{EbelForFil}. 
In general case the PMF of a system with $\tilde{N}$ particles is the potential that 
gives the average force over all the configurations of all the $n+1,\dots,\tilde{N}$ particles acting on a particle j 	at any fixed configuration keeping fixed a set of particles $1, \dots, n$ (see Eq.~(\ref{Z-def})) \cite{Kirkwood1,Kirkwood2} :  
\begin{widetext} 
\begin{eqnarray}\label{avf}
-\nabla_j w^{(n)}(x^{(n)})=\frac{\sum_{\sigma} \int	\rm  dp \rm  dx^{(\tilde{N}-n)}\; \rm d\mu Q \;\nabla_jW^H(p,x,Q)}
{\sum_{\sigma} \int	\rm  dp \rm  dx^{(\tilde{N})}\; \rm d\mu Q \;W^H(p,x,Q)}, j=1, \dots,n
\end{eqnarray}
\end{widetext} 
Above expression $-\nabla_j w^{(n)}$ is the averaged force, i.e. "mean force" on particle j and $w^{(n)}$ is the so-called potential of mean force. 
According to the Ref.~\cite{Chandler} the pair distribution functions can be expressed through the potential of mean force $w^{(2)}$: 
\begin{eqnarray}
\label{crg}
\rm g(r)=\exp (-w^{(2)}(r)/T). 
\end{eqnarray}

As it follows from the left plot of Fig.~\ref{crt4}, the logarithm of PDF, describing the PMF, can be approximated by a 	linear functions at distances smaller than the half of average inter quasiparticle distance with good accuracy. 
Thus the PMFs are almost linear functions of the interparticle distances, which are formed by the contributions of the color Coulomb potentials (strictly speaking color Kelbg pseudopotentials).
Below we will see that at large distances all PDFs (except ${\rm g}_{gg}(r)$) can be described by exponentially screening Yukawa-type effective potential, similarly with electromagnetic plasma.

In lattice QCD calcualtions \cite{Maezawa} the effective Debye mass and running coupling constant at temperatures above $T_c$ have been estimated from the appropriate fit range for free energies $w^{(2)}_M(r,T)$ by the screened Coulomb potential:
\begin{eqnarray} 
\label{dbm}
&&w^{(2)}_M(r,T)/T=V_M(r,T)
\nonumber\\&=&
C(M)\frac{\alpha_{eff}(T,M)}{r}\exp(-m_D(T,M)r),
\end{eqnarray}
where $\alpha_{eff}(T,M)$, $m_D(T,M)$ and $C(M)$ 
are the running coupling constant, the Debye screening mass and the Casimir factors 
for related color channel M respectively 
($C(1)=-\frac{4}{3},C(8)=\frac{1}{6},C(6)=\frac{1}{3},C(3)=-\frac{2}{3}$). 

\begin{figure}[htb]
  \begin{minipage}[ht]{1.0\linewidth}	
	\includegraphics[width=5.5cm,clip=true]{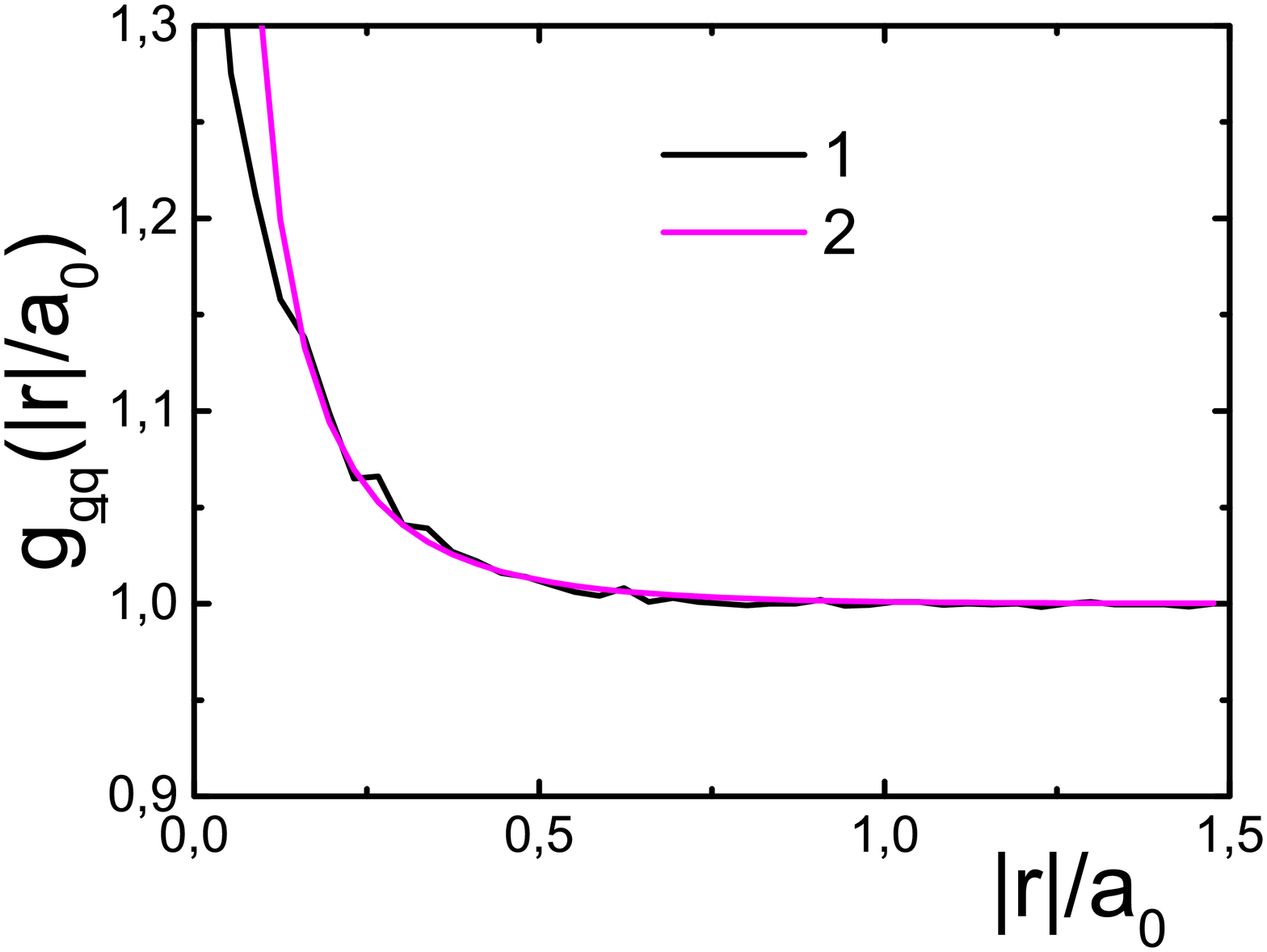}   \\(a)
  \end{minipage} \\
  \begin{minipage}[ht]{1.0\linewidth}  
	\includegraphics[width=5.5cm,clip=true]{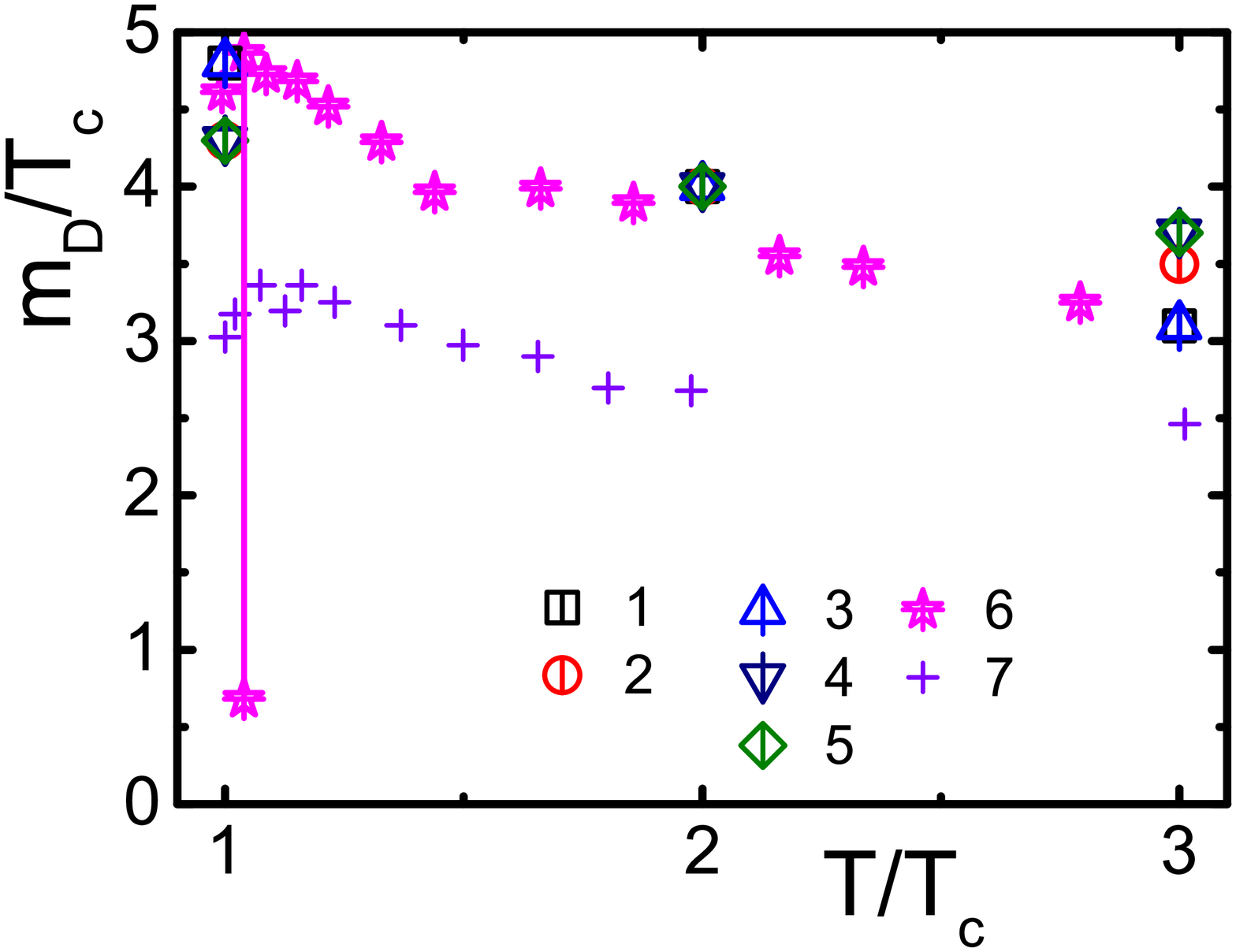}	\\(b)
  \end{minipage} \\		
  \begin{minipage}[ht]{1.0\linewidth}    
	\includegraphics[width=5.5cm,clip=true]{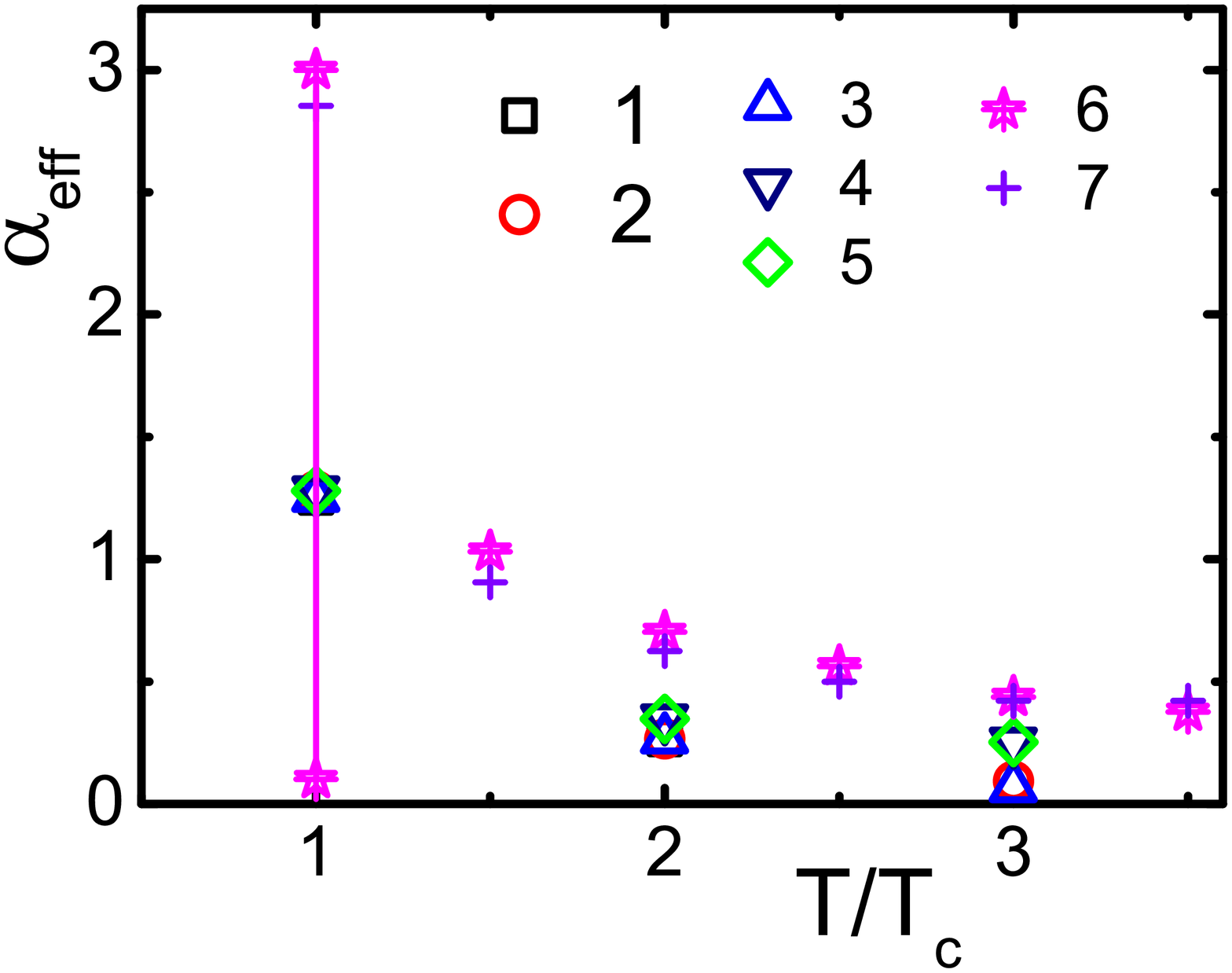}  \\(c)	
  \end{minipage} \\				
	\caption{(Color online) 
(Plot (a)) The pair quasiparticle quark-antiquark distribution function 
at temperature $T=3T_c$    
 and $r_s=0.4$~fm (baryon chemical potential equal to zero, $a_0=1.16$~fm ).		 
Lines: 1~-~ PIMC results; 2~-~ Debye approximation.   
(Plot (b)) The PIMC and lattice QCD \cite{Maezawa} Debye screening masses. PIMC scatters: 1 ~-~ quark--quark;  2 ~-~ antiquark--antiquark; 
3 ~-~ quark--antiquark; 4 ~-~ quark--gluon; 5 ~-~ antiquark--gluon. 
Lattice QCD scatters: 6 ~-~ Wilson quark action; 7 ~-~ staggered quark action.  
(Plot (c)) The running coupling constant. The same notation with the central plot. The PIMC statistical errors 
are of order sizs of the of the related scatters. 
}
	\label{crt31}
\end{figure}

The PIMC simulations allow also to estimate the Debye screening masses $m^{ab}_D(T)$ and 
running coupling constant $\alpha_{eff}(T,M)$ for the quasiparticle pairs free from 
forming bound states (see above).  
According to the equations (\ref{crg}, \ref{dbm}) we have:  
\begin{eqnarray}
\label{dbmm}
&&-\ln (\rm g_{ab}(r))=w^{(2)}_{ab}(r,T)/T=V_{ab}(r,T)
\nonumber\\&=&
\frac{\tilde{\alpha}^{ab}_{eff}(T)}{r}\exp(-m^{ab}_D(T)r), 
\end{eqnarray} 
where 
$\tilde{\alpha}^{ab}_{eff}(T)=<C(M)>\alpha^{ab}_{eff}(T)$ with the averaged Casimir factors $<C(M)> \approx (C(1)+C(8)+C(6)+C(3))/4$. 
The Debye screening masses $m^{ab}_D(T)$ and running coupling constant $\tilde{\alpha}^{ab}_{eff}(T)$ can be estimated from the long distance behavior of the $\rm g_{ab}(r)$, approximated accordingly to (\ref{dbmm}) with the Debye potential $V_{ab}(r,T)$ (see top plots of Fig.~\ref{crt4} and Fig.~\ref{crt31}).
Disagreement with Debye approximation has place only at short interparticle distances, where the main contribution to $\rm \rm g_{ab}	(r)$ is given by the interaction with the nearest neighbor quasiparticle.

The Debye screening masses and running coupling constant, obtained by PIMC and lattice QCD calculations \cite{Maezawa}  correspondingly, are presented on the central and bottom plots of Fig.~\ref{crt31}.
Vertical line shows the scatter region of the Debye masses for color channel 	$M=1, 8, 6, 3$ at $T/T_c=1$, while at other temperatures these regions are approximately of the twice size of the scatters presenting data \cite{Maezawa}.   
Comparison of these results shows quite good agreement between PIMC and lattice QCD results. 
Note that the Debye screening masses, obtained from PIMC,  are closer to ones obtained by Wilson quark action than ones,  obtained from improved staggered quark action. 
Of course, existing disagreements should be further investigated by more detailed PIMC simulations and as well as at smaller lattice spacing.

\section{Pair color  correlation functions} 
Independent estimations of screening parameters can be carried out from pair color correlation functions.
To introduce the pair color correlation functions, let us consider the probability of an elementary configuration for $\tilde{N}$ particles  given by the Eq.~(\ref{Z-def}):
\begin{eqnarray}
	\label{crg1}
&&\rm P^{[\tilde{N}]}(x,Q) = 
\frac{\sum_{\sigma} \int \rm  dp  W^H(p,x,Q)}{\sum_{\sigma} \int	\rm  dp \rm  dx\; \rm d\mu Q \;W^H(p,x,Q)}. 
\nonumber\\
\end{eqnarray}
Total number of particles is huge, so that $\rm P^{[\tilde{N}]}$ in itself is not very useful. 
However, one can also obtain the probability of reduced configuration, where the degrees of freedom of $n$ ($n < \tilde{N}$ ) particles $x_1Q_1,\dots,x_ nQ_n$ are fixed, with no constraints on the remaining $\tilde{N}-n$ indices.  
To do this, one has to integrate (\ref{crg1}) over the remaining degrees of freedom. 
Let us consider the probability for quasiparticles to have positions in the color configuration space $x_1,Q_1, \dots,  x_{n_{q}}, Q_{n_q}$ 	by  expression: 
\begin{eqnarray}
	\label{crg2}
&&\rm P^{[n]}(x^{[n]},Q^{[n_q]})
\nonumber\\&=&
\frac{\sum_{\sigma} \int  \rm  dp \rm  dx_{(n+1)}\; \dots \rm d x_{\tilde{N}} 
\int\rm  dx\; \rm d\mu Q \;W^H(p, \sigma)}{\sum_{\sigma} 
\int \rm  dp \rm  dx\; \rm d\mu Q \;W^H(p,x,Q)}, 
\end{eqnarray} 

Generally speaking, the thermodynamic functions $M_n$ in canonical ensemble can be presented as integrals of the next form: 
\begin{widetext}
\begin{eqnarray}
	\label{therm}
&&<M_n>=
\frac{\sum_{\sigma}  \int \rm  dp \rm  dx\; \rm d\mu Q \; M_n(x,Q) \;W^H(p, x, \sigma)}
	{\sum_{\sigma} \int \rm  dp \rm  dx\; \rm d\mu Q \;W^H(p,x,Q)}, 
	\nonumber\\&
	\mbox{where }\qquad
	&M_n(x,{\rm Q})=
	\sum_{1 \le i_1 < i_2 < \dots < i_n \le N} f(x_{i_1},{\rm Q}_{i_1}; \dots ;x_{i_n},{\rm Q}_{i_n}). 
\end{eqnarray}
\end{widetext}
\begin{figure}[htb]
\begin{minipage}[ht]{0.3\linewidth}
	\includegraphics[width=5.5cm,clip=true]{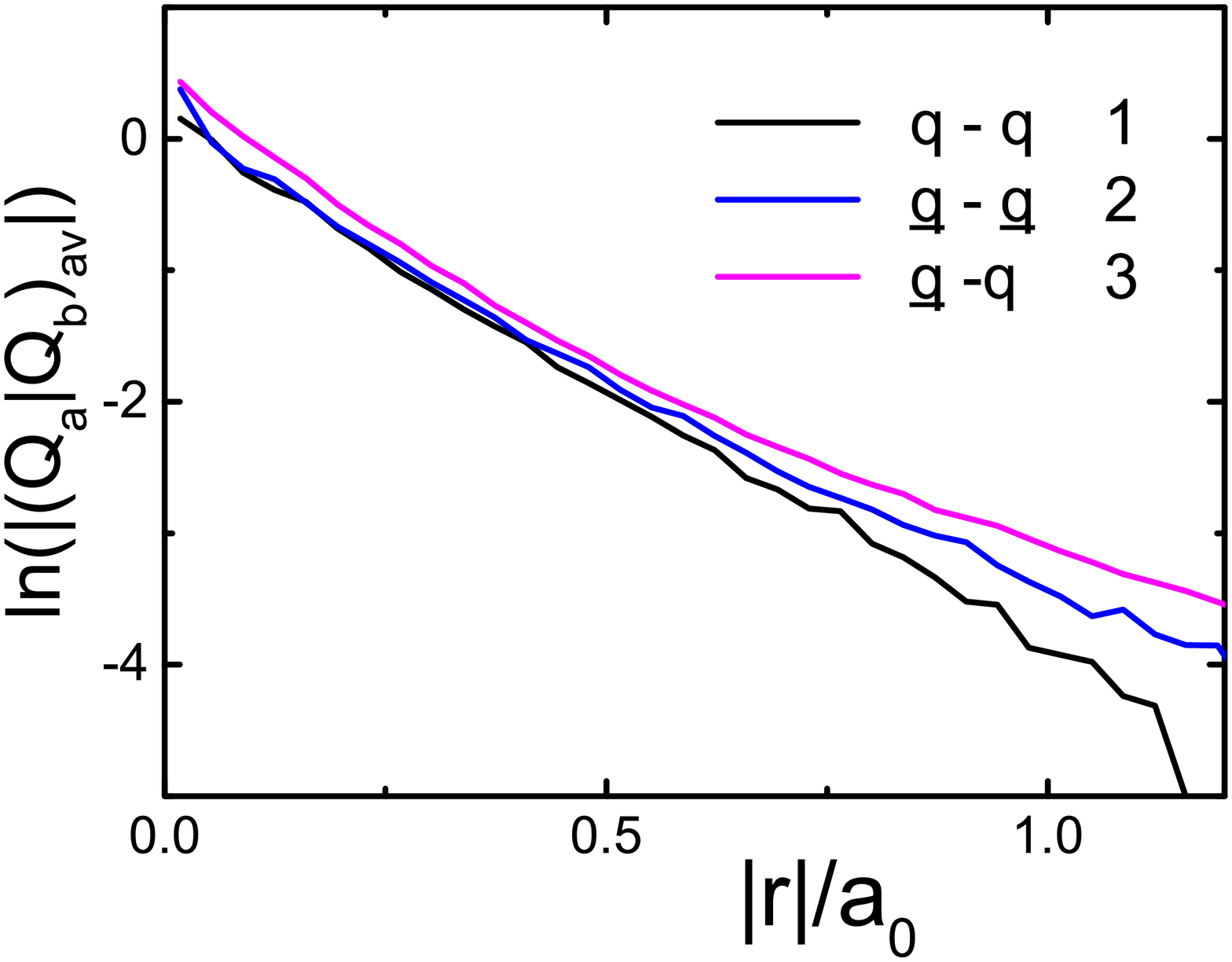} \\(a)
\end{minipage} 
\begin{minipage}[ht]{0.3\linewidth}
	\includegraphics[width=5.5cm,clip=true]{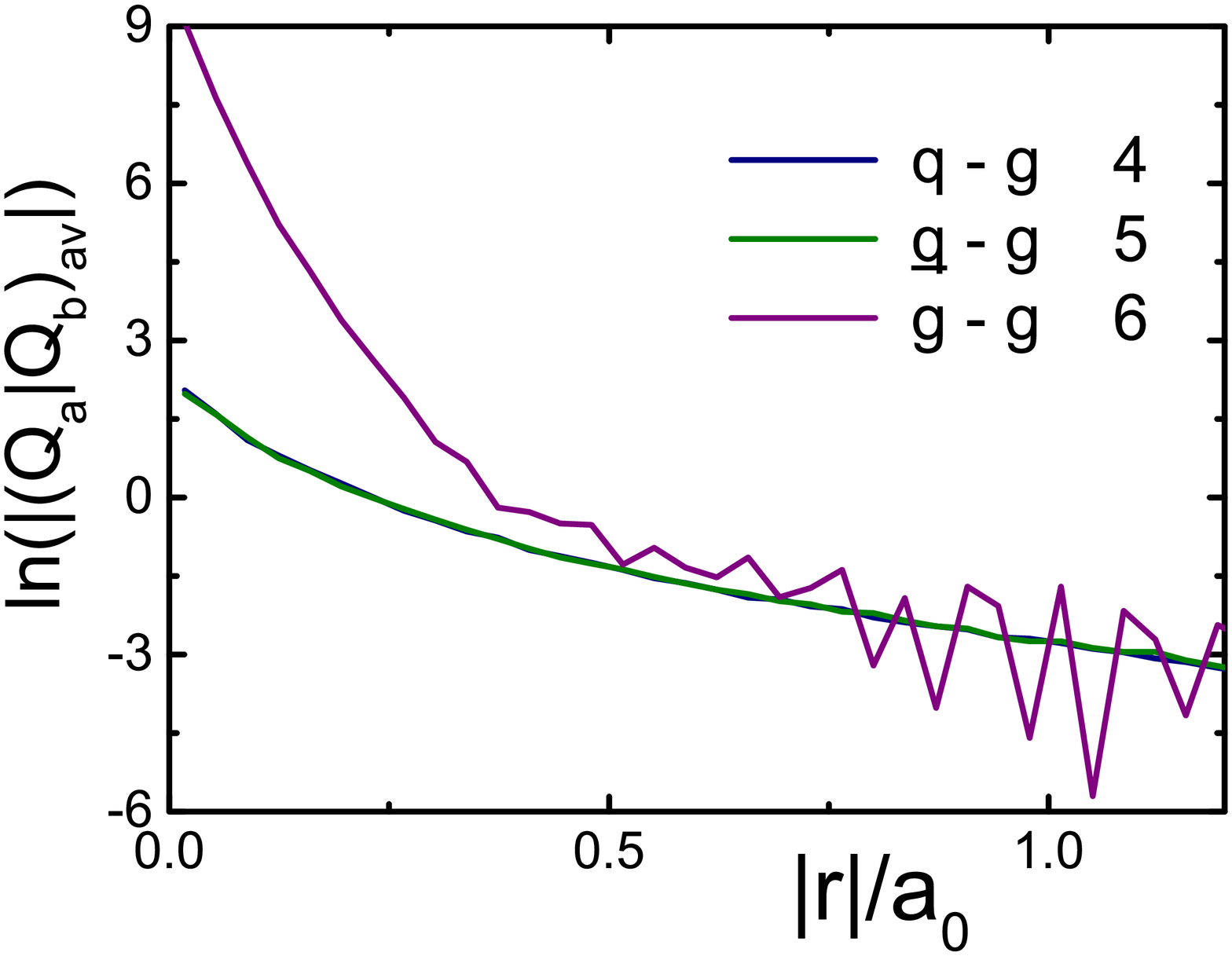}	\\(b)
\end{minipage} 
\begin{minipage}[ht]{0.3\linewidth}
	\includegraphics[width=5.5cm,clip=true]{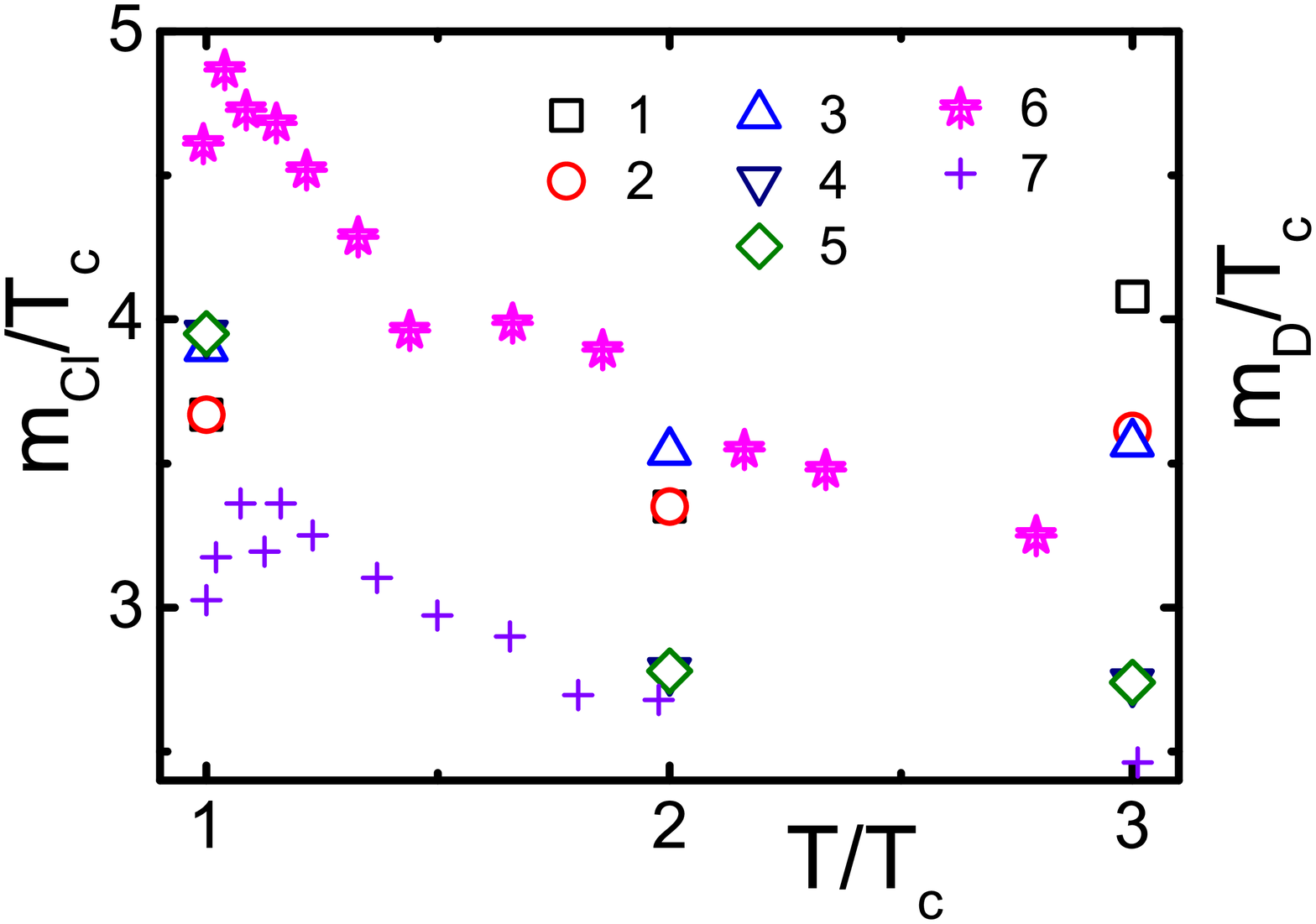}  \\(c)
\end{minipage} 	
	\caption{(Color online) 
The PIMC quark-quark  (Plot (a)) and quark/gluon-gluon (Plot (b))  color  pair correlation functions 
for strongly coupled QGP. 
Lines: 1 ~-~ quark--quark;  2 ~-~ antiquark--antiquark; 3 ~-~ quark--antiquark; 
4 ~-~ quark--gluon; 5 ~-~ antiquark--gluon; 6 ~-~ gluon--gluon.  
Temperature $T=3 T_c$,  
$r_s=0.4$~fm and baryon chemical potential equal to zero ($a_0=1.16$~fm).  
(Plot (c)) Comparison color screening parameter $m_{Cl}/T$ from the color pair correlation functions 
and Debye screening mass \cite{Maezawa}. Scatteres: 1 ~ quark - quark;  2 ~ antiquark - antiquark; 3 ~ quark - antiquark; 
4 ~ quark - gluon; 5 ~ antiquark - gluon. Lattice QCD  \cite{Maezawa}: 
6 ~-~ Wilson quark action; 7 ~-~ staggered quark action. 
	}
	\label{clms}
\end{figure} 

Thus, for example, for $n_q$ quasiparticles the number of terms in Eq.~(\ref{therm}) is equal to $N_q!/(N_q-n_q)!n_q!$: 
\begin{eqnarray}
	\label{therm1}
&&<M_n> =\frac{N_q!}{(N_q-n_q)!n_q!}
\nonumber\\&\times&
\int  \rm  dx_1\; \rm d\mu Q_1 \dots \rm  dx_{n_q}\; \rm d\mu Q_{n_q} 
\nonumber\\&\times&
	f(x_1,{\rm Q}_1; \dots ;x_{n_q},{\rm Q}_{n_q} )  P^{[n]}(x^{[n_q]},{\rm Q}^{[n_q]}).
\end{eqnarray} 

Let us introduce the dimensionless color correlation functions by the next equations:
\begin{eqnarray}  
	\label{crg5}
	&&\frac{N^{n_q}}{V^{n_q}} \rm g^{[n_q]}(x^{[n_q]},Q^{[n_q]}) 
\nonumber\\&=&	
	\frac{N_q!}{(N_q-n_q)!n_q!}
	P^{[n_q]}(x^{[n_q]},Q^{[n_q]}).   
\end{eqnarray} 
For example, pair quark-quark color correlation function  can be defined through scalar product of 8D color Wong vectors:
\begin{eqnarray}
\label{clf}
&&\tilde{M}_2(r,x,{\rm Q})
\nonumber\\&=&
\sum_{1 \le i_1 < i_2 \le N} \delta(r-|x_{1_{i_1}}-x_{i_2}|) ({\rm Q}_{i_1} \cdot {\rm Q}_{i_1}). 
\end{eqnarray}
Thus, the averaged over canonical ensemble pair color correlation function looks like:
\begin{widetext}
\begin{eqnarray}
	\label{crg7}
({\rm Q}_{a} \cdot {\rm Q}_{b})_{\rm av}(r)=\frac{N_q^2}{V^2}\int \rm  dx_{1,a} \rm  dx_{2,b}  \rm  d\mu {\rm Q}_{1,a} \rm  d\mu {\rm Q}_{2,b} 
	\delta(r-|x_{1,a}-x_{2,b}|)({\rm Q}_{1,a} \cdot {\rm Q}_{2,b})\rm \rm g_{ab}(x_{1,a},Q_{1,a}; x_{2,b},Q_{2,b}), 
\end{eqnarray}
\end{widetext}
where a, b  correspond to quarks and $\rm g^{[2]}(x^{[2]},Q^{[2]})$ is the pair color distribution function of the system. 
Definition of color pair distribution functions for quasipaticles of different types (a and b correspond to quark, antiquark or gluon) is the similar.
These functions are negative due to contribution of quasiparticle attraction and can be approximated by decaying negative exponents:
\begin{eqnarray}
\label{clmm}
({\rm Q}_{a} \cdot {\rm Q}_{b})_{\rm av}(r) \approx -\exp(-m^{ab}_{Cl}(T)r).
\end{eqnarray} 
The logarithms of module of the pair color distribution function are shown on the left and central plots of Fig.~\ref{clms}.
Thus, according to its definition, the color screening mass $m^{ab}_{Cl}(T)$ can be estimated from the slope of this logarithm, approximated by straight line at distances less than the average interparticle distance $r_s=0.4$~fm.
 
Analysis of Fig.~\ref{clms} confirms exponential decay of color pair correlation functions and allows us to calculate the color screening masses in strongly coupled QGP, which turn out to be of order the Debye screening mass, obtained in PIMC and lattice QCD \cite{Maezawa} calculations. 
Scatter of the values $m^{ab}_{Cl}(T)$ for different types of quasiparticles is also connected with different values of the related Casimir factors.

\section{Quantum ``tails'' in momentum distribution functions}
The momentum distribution function $w_a(|p|)$ can be obtained by integration of Wigner function over all quasiparticle coordinates and momenta except the momentum of some quasiparticle of type a:
\begin{eqnarray}
W_a(p_a) =  
\int	\rm  dx\; \rm  dp\; \rm d\mu Q \; \delta(p_a-p) W^H(p,x,Q)
\label{wdp}
\end{eqnarray}
where $a=q,\overline{q},g$. 
From physical point of view the momentum distribution function $w_a(|p|)$ gives probability density for quasiparticle to have the momentum $p$.   

Non-ideal classical systems of particles are described by Maxwell distribution (MD) (proportional to $\exp(-(p\lambda_a)^2/4\pi\hbar^2)$ )) even when coupling is strong, due to the commutativity of the kinetic and potential energies.
Ideal quantum systems of particles, due to the quantum statistics , are described by Fermi or Bose momentum distribution functions.
Interparticle interaction may be able to cause formation of the bound states of two particle or many particle clusters and, hence, influence on the momentum distribution function.
Moreover, interaction of not-bounded quantum particle with its surroundings may restrict the available volume of configuration space for the particle and, due to the uncertainty principle, can also influence on the momentum distribution function \cite{Galitskii,Kimball,StarPhys,Eletskii,Emelianov,Kochetov,StMax}. 
Thus, all these physical factors can modify the momentum distribution function, making it non-Maxwellial at high momenta.

One of the main objectives of this paper is to study the influence of strong interaction between quasiparticles on the color quasiparticle momentum distribution functions in QGP. 

It has been shown in \cite{Galitskii,Kimball,StarPhys,Eletskii,Emelianov,Kochetov,StMax}, that in fully ionized electromagnetic plasma the momentum distribution function at high momenta can be described by the sum of the MD and the product of $const/p^{8}$ and the Maxwell distributions with effective temperature that exceeds the temperature of medium (short notation – P8)\cite{StMax}.

\begin{figure}[htb]
 \begin{minipage}[ht]{1.0\linewidth}
	\includegraphics[width=5.5cm,clip=true]{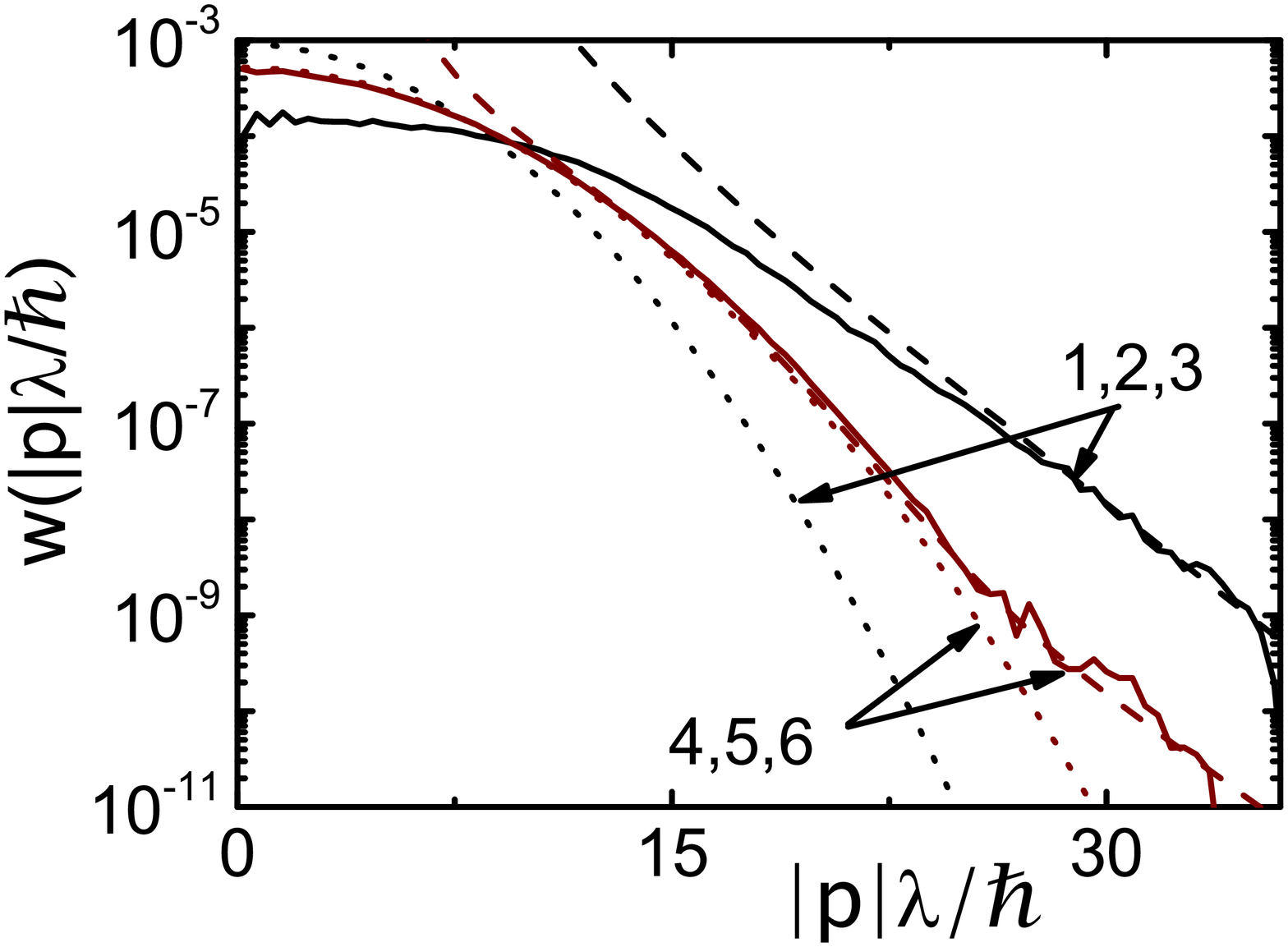}  \\(a)
 \end{minipage}	\\	
 \begin{minipage}[ht]{1.0\linewidth} 
	\includegraphics[width=5.5cm,clip=true]{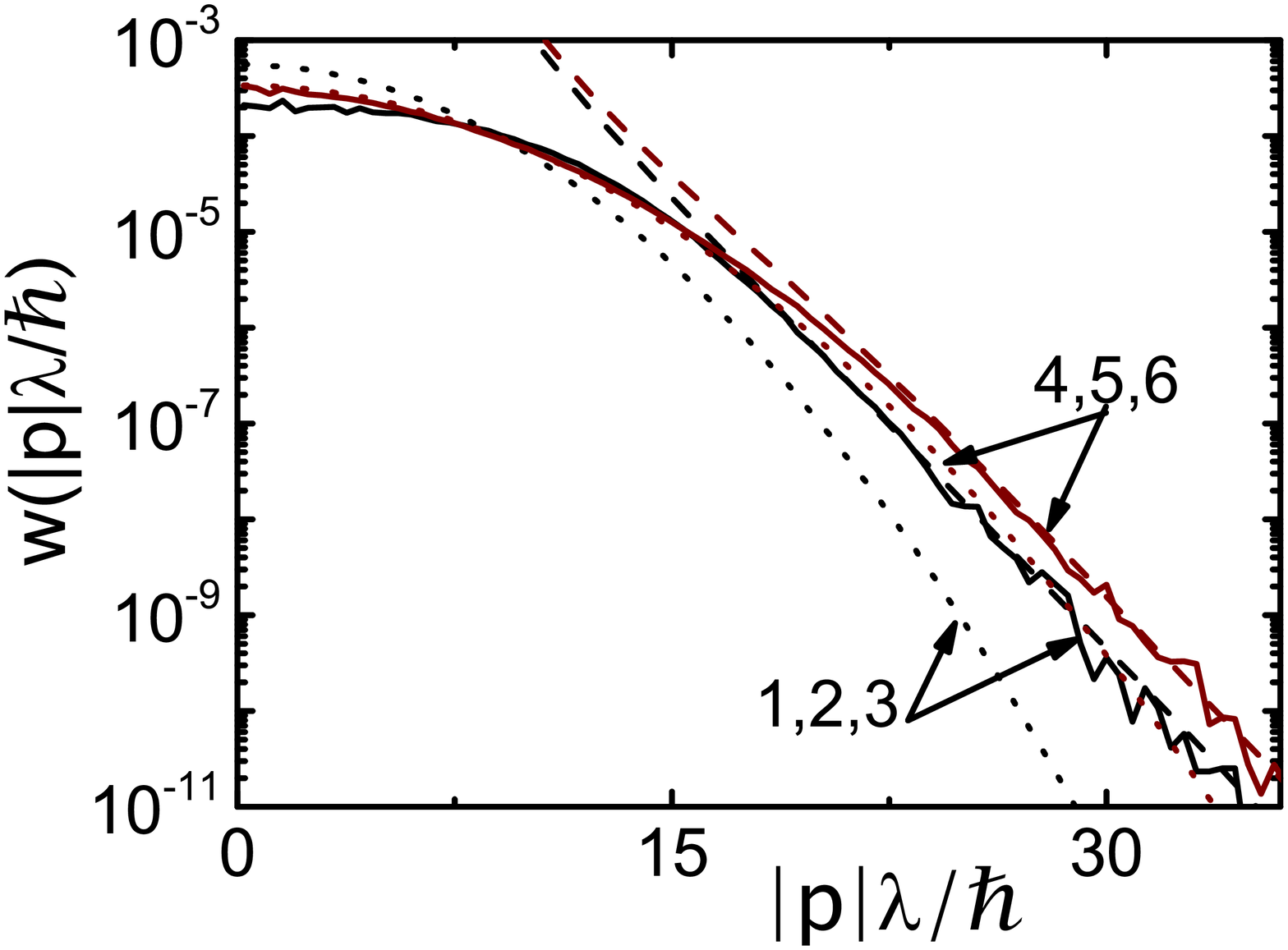}   \\(b)
 \end{minipage}	\\		
 \begin{minipage}[ht]{1.0\linewidth}  
	\includegraphics[width=5.5cm,clip=true]{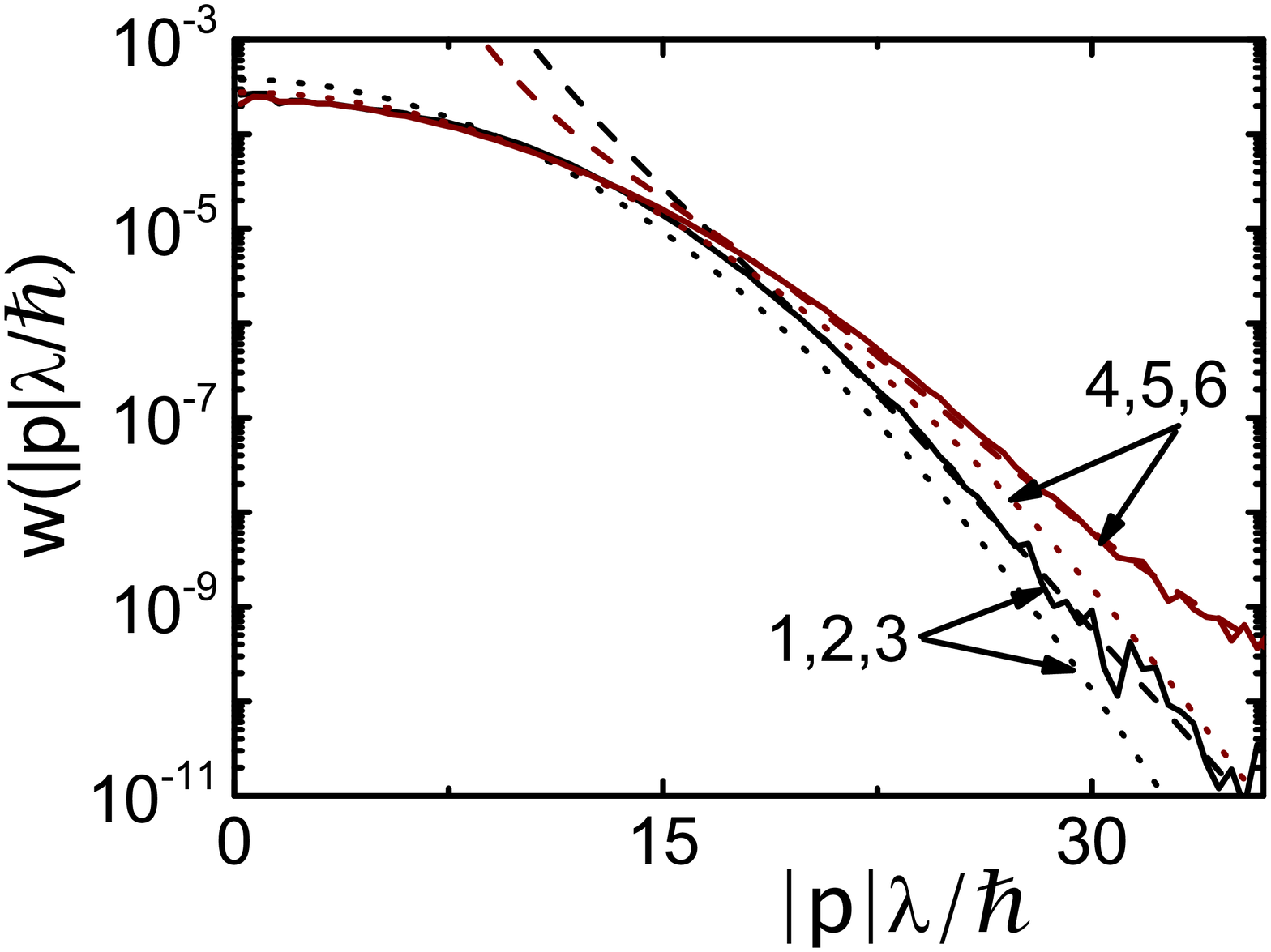}	 \\(c)
 \end{minipage}	\\			
	\caption{(Color online)   
		The averaged over color, flavor and spin variables momentum distribution functions 
		$w_a(|p|) \, (a=q, g)$ for quarks and gluons of the strongly coupled QGP plasma at $r_s=0.4$~fm. 
		(Plot (a)) Temperature $T/T_c=1$. Lines: 1(4) ~-~ the Maxwell distribution for quarks(gluons); 
		2(5) ~-~ the PIMC results for for quarks(gluons); 3(6) ~-~ the high momentum asymptotics (P8) \cite{StMax}.  
		(Plot (b)) Temperature $T/T_c=2$. The same notation with left plot. 
		(Plot (c))  Temperature $T/T_c=3$. The same notation with central plot.
		The momentum distribution functions for quarks and antiquarks practically coincide with each other,  
		here $\lambda$ is the thermal wave length and $\hbar$ is the Plank's constant. 
		All momentum distribution functions are normalized to unity.		
		Oscillations of  the PIMC distribution functions designate statistical errors. 
	}
	\label{wpt4}
\end{figure}

The results of PIMC calculations of quark and gluon momentum distribution functions are shown in Fig.~\ref{wpt4}. 
Fig.~\ref{wpt4} shows also two related dependencies: Maxwell distributions (lines 1 and 4) 
and analytical high momentum asymptotic (P8) (lines 4 and 6). 
Here the constant and the effective temperature in P8 have been considered as adjustable parameters, set to fit PIMC momentum distribution functions at high momenta.
As it follows from the analysis of Fig.~\ref{wpt4}, the dependences P8 can reliably fit the PIMC distributions at high momenta and confirm appearance of quantum ``tails''.

Trends in behavior of momentum distribution functions, when temperature increases, can be understood from comparison of results for temperatures $T/T_c=1$ and $T/T_c=3$, while  $r_s=0.4$ is fixed (see plots of Fig.~\ref{wpt4}).  

The basic physical reason of difference between behaviors of momentum distribution functions of quarks and gluons is that the quadratic Casimir value for qluons, responsible for interparticle interaction, is significantly larger than the one for quarks.

%

\section{Conclusion}	
Thermodynamic and kinetic properties are important for theoretical description of equilibrium states of quark-gluon plasma and need to be considered within unified appropriate model. 
In the framework of considered constituent quasiparticle model of quark-gluon plasma (QGP) the matrix elements of density operator and the Wigner function in the color phase space are presented in form of color path integrals with Wiener and SU(3) group Haar measures.  
The obtained explicit expression of the Wigner function 
resembles the Maxwell--Boltzmann distribution on momentum variables, 	but with quantum corrections.
This approximation contains also oscillatory multiplier describing quantum interference between coordinates and momenta. 

Monte Carlo calculations of quark and gluon densities, momentum and spatial pair distribution functions for strongly coupled QGP in thermal equilibrium at zero baryon chemical potential have been carried out.
The Debye screening mass and running coupling constant have been obtained from the spatial pair distribution function; the results are in agreement with available lattice data. 
Gluon bound states in form of glueballs has been found at temperatures of order $T=3T_c$ and densities corresponding to the average interparticle distance $r_s\ge 0.4$~fm.

Comparison with classical Maxwell–Boltzmann distribution shows the significant influence of interparticle interaction on high energy asymptotics of the momentum distribution functions, resulting in appearance of quantum ``tails''. 
New pair color correlation function, color distribution functions and color screening mass have been developed and discussed.
Quantum effects have proven to be of primary importance in these simulations, showing a valuable understanding of the internal structure of QGP.

Our analysis is still too simplified and incomplete. 
It is still confined only to the case of zero baryon chemical potential. 
The input data of the model also requires refinement.
Work on these problems is in progress. 
	
\section{Acknowledgements}
We acknowledge stimulating discussions with 
Profs.	A.N.~Starostin, D.~Blaschke and Yu.B.~Ivanov.
Authors acknowledge the program of fundamental research of the Presidium
of the Russian Academy of Sciences ``Condensed matter and plasma at high energy densitie'' for financial support.

\bibliographystyle {apsrev}
\bibliography{Filinov}

\end{document}